%% file: hbg.tex
\documentclass[useAMS,twocolumn,usenatbib]{mn2e}
\usepackage{amssymb}
\usepackage{amsmath}
\usepackage{graphicx}
\DeclareGraphicsExtensions{.eps, .ps, .eps.gz, ps.gz}
\usepackage[british]{babel} 
\usepackage{placeins}
\usepackage{subfigure}
\usepackage{color}
\usepackage{multirow}
\usepackage[ps2pdf]{hyperref}
\hypersetup{
  pdftitle={Comparing halo bias from abundance and clustering},
  pdfauthor={K. Hoffmann, J. Bel, E. Gaztanaga,}
}

\begin{document}
  

\title{Comparing halo bias from abundance and clustering}
\author[K. Hoffmann, J. Bel, E. Gazta\~naga] 
{K. Hoffmann$^{1}$, J. Bel$^{2}$, E.Gazta\~naga$^{1}$\\
$^{1}$Institut de Ci\`{e}ncies de l'Espai (ICE, IEEC/CSIC), E-08193 Bellaterra (Barcelona), Spain\\
$^{2}$INAF - Osservatorio Astronomico di Brera, Via Brera 28, 20122 Milano, via E. Bianchi 46, 23807 Merate, Italy\\
} 

\date{Received date / Accepted date}

\maketitle

\begin{abstract}
We model the abundance of haloes in the $\sim(3 \ \text{Gpc}/h)^3$ volume of the MICE Grand Challenge
simulation by fitting the universal mass function with an improved Jack-Knife error covariance estimator that
matches theory predictions. We present unifying relations between different fitting models and new
predictions for linear ($b_1$) and non-linear ($c_2$ and $c_3$) halo clustering bias.
Different mass function fits show strong variations in their performance when including 
the low mass range ($M_h \lesssim 3 \ 10^{12}  \ M_{\odot}/h$) in the analysis.
Together with fits from the literature we find an overall variation
in the amplitudes of around $10$\% in the low mass and
up to $50$\% in the high mass (galaxy cluster) range  ($M_h > 10^{14} \ M_{\odot}/h$). These
variations propagate into a $10$\% change in $b_1$ predictions and a $50$\% change in $c_2$ or $c_3$.
Despite these strong variations we find universal relations between $b_1$ and $c_2$ or $c_3$ for which
we provide simple fits. Excluding low mass haloes, different models fitted with reasonable
goodness in this analysis, show percent level agreement in their $b_1$ predictions, but are systematically
$5-10$\% lower than the bias directly measured with two-point halo-mass clustering. This result confirms
previous findings derived from smaller volumes (and smaller masses).
Inaccuracies in the bias predictions lead to $5-10$\% errors in growth measurements. They also affect
any HOD fitting or (cluster) mass calibration from clustering measurements.
\end{abstract}
 

\input{sections/introduction.tex}

\input{sections/simulation_and_massfunction.tex}

\input{sections/pbs_bias.tex}

\input{sections/bpbs_vs_bxi.tex}

\input{sections/conclusion.tex}


\section*{Acknowledgements}

Funding for this project was partially provided by the Spanish Ministerio de Ciencia e Innovacion (MICINN), project AYA2009-13936,
Consolider-Ingenio CSD2007- 00060, European Commission Marie Curie Initial Training Network CosmoComp (PITN-GA-2009-238356) and research
project 2009- SGR-1398  from Generalitat de Catalunya.
JB acknowledges useful discussions with Emiliano Sefusatti and support of the European Research
Council through the Darklight ERC Advanced Research Grant (\#291521).
KH is supported by beca FI from Generalitat de Catalunya. He also acknowledges the Centro de Ciencias de Benasque Pedro Pascual
where parts of the analysis were done.
The MICE simulations have been developed by the MICE collaboration at the MareNostrum
supercomputer (BSC-CNS) thanks to grants AECT-2006-2-0011 through AECT-2010-1-0007.
Data products have been stored at the Port d'Informació Científica (PIC).

We thank Martin Crocce, Pablo Fosalba, Francisco Castander, Fabien Lacasa and Shun Saito
for interesting and useful comments.

\bibliographystyle{mnbst}
\bibliography{hbg.bib}

\appendix

\input{sections/appendix.tex}

\end{document}

%% file: sections/introduction.tex
\section{Introduction}\label{sec:introduction}

      Observations of structures in the large-scale distribution of galaxies are a powerful
      tool for constraining cosmological models. However, such constraints require
      a model which connects the galaxy distribution to the matter density field.
     Various observations support the gravitational instability paradigm in which galaxies form
     in potential wells, generated by the gravitational collapse of dark matter into haloes.
     The relation between the halo and full matter density fields ($\rho_h$ and $\rho_m$
     respectively) is therefore a crucial ingredient for a precise large-scale structure analysis.
     In fact,  the uncertainties in this relation strongly increase the errors in the Dark Energy 
     equation of state or gravitational growth index from future galaxy surveys
     \citep[e.g.][]{Eriksen2015}.
     
     A formal approach for describing the halo-matter density relation was suggested by \citet{FG} as the Taylor
     expansion around the matter density contrast, $\delta_m$, at the same position, known as bias function
    \begin{equation} 
    \delta_h({\bf r})=F[\delta_{m}({\bf r})] \simeq \sum_{i=0}^{N}\frac{b_i}{i!}\delta_{m}^i({\bf r}),
    \label{eq:biasfunction}
    \end{equation}
    where $\delta({\bf r}) \equiv (\rho({\bf r})- \bar{\rho})/\bar{\rho}$, $\bar{\rho}$ 
    is the mean density and ${\bf r}$ denotes the spatial position. For the construction of
    $\delta(\bf r)$ the density field 
    is commonly smoothed with a top-hat filter of size $R$.
    The coefficients $b_i$ are the so called
    bias parameters, while we will investigate non-linear bias in terms of the ratios
    $c_2\equiv b_2/b_1$ and $c_3\equiv b_3/b_1$.
    The relation in equation (\ref{eq:biasfunction}) corresponds to a local bias model in which the
    density of galaxies is fully determined by the matter density at the same position, while
    environmental effects are not considered.
    Recent studies demonstrated that the local model is inadequate as tidal forces of
    the surrounding large-scale structure generate non-local contributions to the bias function
    \citep[e.g.][]{chan12, Baldauf2012}. However, the two-point correlation, commonly used to study galaxy
    clustering, is primarily sensitive to the linear bias parameter $b_1$ at scales between $20-60$
    $h^{-1}$Mpc.	 Due to the level of precision achieved in our analysis we will not take non-linear and
    non-local bias contributions to the two-point correlation into account. Note that, especially at
    smaller scales, such a negligence would not be appropriate \citep[e.g.][]{Saito2014, Biagetti14}.
  
     Besides the clustering also the abundance of haloes as a function of halo mass
    (known as the mass function) is related to the bias function.
    This relation can be understood with the peak-background
    split model \citep[hereafter referred to as PBS model,][]{bardeen1986, cole1989, mo96}.
    In this model, large-scale density fluctuations are superposed with fluctuations 
    at small scales.
    These large-scale density fluctuations modulate the background cosmology
      (i.e. the mean density and the Hubble rate) around small-scale fluctuations \citep[e.g.][]{Martino09}.
      The critical density contrast for gravitational collapse therefore depends on the environment.
      In regions with large-scale overdensities more small-scale fluctuations collapse to haloes than
      in underdense regions. This effect modifies the abundance of haloes and also their spatial
      distribution as they follow the pattern of the peaks of large-scale fluctuations. Haloes are therefore
      biased tracers of large-scale fluctuations in the full matter density field. For a given matter power spectrum
      the halo bias parameters can be predicted from the mass function via the PBS model.
    
   The PBS bias predictions can be used to determine the dark matter clustering
   from observed galaxy distributions if the halo masses of a given tracer sample are known
   (or the other way round).
      Such an analysis requires that the bias parameters, predicted from the mass function, are
   equivalent with the bias which affects the clustering. Studies of this equivalence have revealed that the PBS 
   predictions for the linear bias $b_1$ are around $10\%$ below measurements from two-point
   clustering statistics.
   Such deviations might result from assumptions of the PBS model, such as 
   spherical collapse, or a local bias relation \citep[e.g.][]{Mo97, Desjacques10, Paranjape13, Schmidt13}.
   Further numerical effects, like the definition of haloes in N-body simulations, or systematic effects such as 
   the parametrisation and fitting procedure of the mass function might 
   contribute to the discrepancy between the bias from PBS and clustering \citep[e.g.][]{HuKra03, MS&S10}.
    Predictions of the PBS for the relation between halo mass and bias are also employed
    in Halo Occupation Distribution models to predict the bias
    as a function of galaxy properties, such as luminosity or color \citep[e.g.][]{CooraySheth02, MoKraDaGott11, Coupon12, Carretero15}.
    Inaccuracies of the PBS can affect such halo model predictions for galaxy bias or the average number
    of galaxies per halo. Moreover, haloes of equal mass could have different galaxy occupation,
    depending on their environment \citep[e.g.][]{Pujol14}. 
    Besides clustering analysis the PBS can be employed for estimating the 
    lower mass threshold (or mass-observable relation) of observed galaxy samples. This so-called
    self calibration method \citep{LimaHu04, LimaHu05} uses the fact that both, the 
    clustering and the abundance of haloes, depend on halo mass. Inaccuracies
    of the PBS model can change the estimation of halo mass thresholds and therefore change
    the cosmological parameters inferred from such an analysis \citep[e.g.][]{Wu2010,M&G11}.
    
    The broad application of the PBS model in large-scale structure analysis and the 
    precision of abundance and clustering measurements from incoming
    observational data calls for a detailed validation of the PBS bias predictions.
    The purpose of this analysis is to pursue the study of deviations between
    halo bias measurements from clustering and PBS predictions using
    the wide mass range of the MICE Grand Challenge (hereafter referred to as MICE-GC)
    simulation \citep{mice1, mice3, mice2, Carretero15, paper1}. We thereby focus
    on the effect of mass function parametrisation and fitting on PBS bias predictions.
    The mass function fits are affected by the error estimations.
    Our analysis therefore includes a detailed study of the mass function error and 
    covariance which leads us to an improvement of the standard 
    Jack-Knife estimator.
    The study of PBS bias predictions includes non-linear bias parameters 
    which are important for an analysis of higher-order correlations of the large-scale halo 
    distribution and two-point correlations at small scales.
    We further compare the mass function fits and bias predictions with results from 
    the literature based on different simulations, to verify a universal behaviour of these
    quantities.
    
    This paper is organised as follows. In Section 2 we present the MICE-GC simulation and mass function fits.
    In Section 3 we present new galaxy bias predictions which we compare with the literature and
    find a universal relation between bias parameters.   In Section 4 we compare these
    predictions with the bias directly measured from
    the two-point halo-matter cross-correlations of the MICE-GC simulation. The comparison with higher-order clustering will
    be presented in a separate paper (Bel, Hoffmann \& Gazta\~naga, in preparation). A summary is given together with
    our conclusions in Section 5.
    In Appendix B we present a new method to improve the Jack-Knife
    covariance matrix estimation. This method can also be easily generalised to other statistics, such
    as the two-point correlation function (Hoffmann et al., in preparation).

%% file: sections/simulation_and_massfunction.tex
    \section{Simulation \& Halo Mass Function}
     
      Our analysis is based on dark matter haloes, identified in the  comoving outputs of the
      MICE-GC simulation at the redshifts $z = 0.0$ and $0.5$.
      Starting from small initial density fluctuations at redshift $z=100$ the formation of large-scale
      cosmic structure was computed with $4096^3$ gravitationally interacting collisionless particles
      in a $3072$ $h^{-1}$Mpc box using the GADGET - 2 code \citep{springel05} with a softening
      length of $50$ $h^{-1}$kpc. The initial conditions were generated using the Zel'dovich
      approximation and a CAMB power spectrum with the power law index of $n_s = 0.95$, which
      was normalised to be $\sigma_8 = 0.8$ at $z=0.0$.  The cosmic expansion is described by the
      $\Lambda$CDM model for a flat universe with a mass density of
      $\Omega_m$ = $\Omega_{dm} + \Omega_b = 0.25$. The density of the
      baryonic mass is set to $\Omega_b = 0.044$ and $\Omega_{dm}$ is the dark matter density.
      The dimensionless Hubble parameter is set to $h = 0.7$. More details and validation tests on this
      simulation can be found in \citet{mice1}.
      
      Dark matter haloes were identified as Friends-of-Friends groups \citep[hereafter referred to as
      FoF groups,][]{davis85} with a redshift independent linking length of $0.2$ in units of the
      mean particle separation. These halo catalogues and the corresponding validation checks are
      presented in \citet{mice2}.
      To study the galaxy bias as a function of halo mass  we divide the haloes
      into the four redshift independent mass samples M0, M1, M2 and M3, shown in
      Table \ref{table:halo_samples}. These samples span a mass range from Milky Way like haloes up to
      massive galaxy clusters. In our analysis we consider mass function fits over different mass ranges
      which we label as M0123, M123, M23 or M012, following the notation in Table \ref{table:mass_ranges}.
 
      \begin{table}
        \begin{tabular}{c  c c c }
       sample  & mass range [$10^{12} h^{-1}M_{\odot}$]   &  $N_p$ & $N_{h}$\\
           \hline 
             M0	&	$0.58 - 2.32$	&	$20-80$	     &  $122300728$ \\
             M1	&	$2.32 - 9.26$	&	$80-316$        &  $31765907$ \\
             M2	&	$9.26 - 100$	&	$316-3416$    &  $8505326$ \\
             M3	&	$\ge 100$	      &	$\ge3416$ 	    &   $280837$ \\
         \end{tabular}      \centering
        \caption{Halo mass samples. $N_p$ is the number of dark matter particles per halo, $N_{h}$ is the
        number of haloes per sample in the comoving output at redshift $z=0.5$.}
        \label{table:halo_samples}
      \end{table}

      \subsection{Mass function definition and measurement}
      The unconditional mass function, $dn(m)$, is defined as the comoving number density of haloes with
      masses between $m$ and $m+dm$.
	The mass function can be written in a form which is nearly independent of redshift,
      	cosmology and initial power spectrum \citep{PS74, bond91, Sheth99} as
      	
      \begin{equation}
      \nu f(\nu) \equiv   \frac{m}{\bar \rho}\frac{dn(m)}{d\ln \ \nu},
      \label{eq:massfunction_def}
      \end{equation}
      where $\bar \rho$ is the mean comoving mass density.
      The height of density peaks is defined as
       \begin{equation}
          \nu \equiv \delta_{c}^2 / \sigma_m^2(m),
       \end{equation}
      where  $\delta_{c} = 1.686$ is the critical density for spherical collapse
      (which is the exact solution value for the spherical collapse in an
        Einstein-de Sitter universe). The variance of matter density fluctuations, 
        $\sigma_m^2(m)$, smoothed with a spherical top-hat window with radius
        $R(m) = (3 m / 4 \pi\rho)^{1/3}$, can be calculated as
      \begin{equation}
      \sigma_m^2(m) = \int \frac{dk}{k} \frac{k^3 P(k)}{2 \pi^2} W^2(kR(m))
      \label{eq:sigma}
      \end{equation}
      where  $W(x) = (3/x^3)(\sin x - x \cos x)$ is the spherical top-hat window in Fourier space
      and $P(k)$ is
      the linear power spectrum. Note that $m$ refers to the matter density field
      when it appears as lower index and to the mass, enclosed by $R(m)$, when
      used as a variable. We measure the mass function in the MICE-GC
      simulation at redshift $z$ and convert it to $\nu f(\nu)$,
      to predict the halo bias parameters $b_1$, $c_2$ and $c_3$ via
      the PBS theory \citep{bardeen1986,cole1989,mo96}.
      We do not apply the halo mass correction suggested
      by \citet{warren06} for low mass resolution, since we analyse haloes  down to $20$ particles,
      while this correction was only proposed for larger numbers of particles per halo. 
      Furthermore, it is not clear that the FoF mass, corrected in such a way is 
      closer to the halo mass on which the PBS model is based on.
       But note that our results do not depend on this correction as
       illustrated in Fig. \ref{fig:massfct_comp}. More details about how we measure the mass function are given in Appendix
      \ref{app:mass_function_measurement}. 

\begin{figure}
      	\centerline{\includegraphics[width=110mm,angle=270]{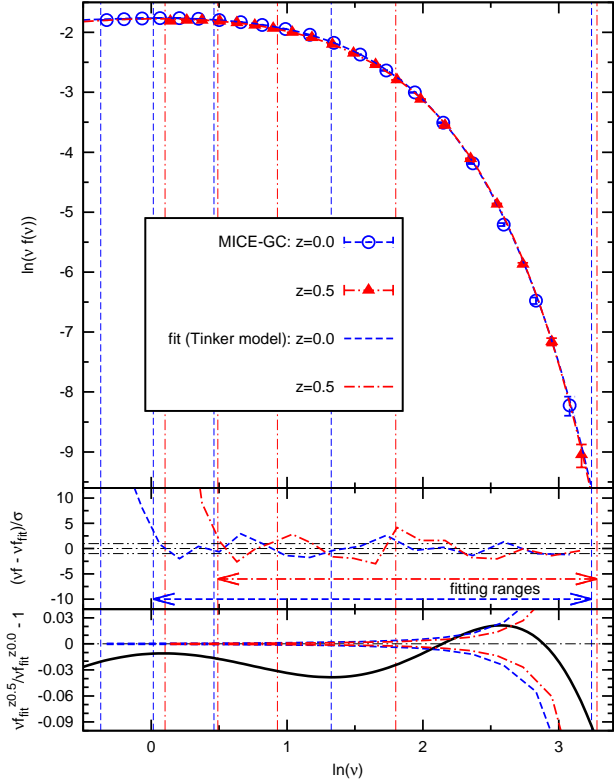}}
      	\caption{{\emph Top:}  unconditional halo mass function, defined in equation
      	(\ref{eq:massfunction_def}), as a function of the peak height $\nu \equiv \delta_{c}^2 / \sigma^2(m)$. Symbols show MICE-GC measurements with $1\sigma$
      	errors based on FoF groups at the redshifts $z=0.0$ and $z=0.5$ (blue circles and red triangles respectively). Lines show 
      	the mass function model of \citet{Tinker10}, fitted to the measurements over the mass range M23
      	in the same colour coding as the symbols.
      	{\emph Center:}  significance of the deviation between measurements and fits.
      	{\emph Bottom:} 
      	relative deviation between the fits at z=0.0 and z=0.5 (black solid line).
      	The $1\sigma$ errors of the measurements are shown as lines in the same colour coding as
      	in the top panels. Vertical blue dashed and red dash-dotted lines denote the limits of the halo
      	mass samples M0-M3 at  z=0.0 and z=0.5 respectively, given in Table \ref{table:halo_samples}.}
      	\label{fig:massfctfit}
      	\end{figure}

      Our measurements of $\nu f(\nu)$ at $z=0.0$ and $z=0.5$ are shown as symbols in
      Fig. \ref{fig:massfctfit}. As expected, they agree visually with
      the idea of a weak redshift dependence
      for FoF mass functions when using a redshift independent linking lengths \citep[e.g.][]{PS74, Jenkins01, MoKraDaGott11}.
       Errors and covariances of the measurements were derived with a new 
      estimator which combines the JK approach with predictions for sampling variance from
      the power spectrum (see Appendix \ref{app:covar}).
      We also show in Fig. \ref{fig:massfctfit} fits to the measurements, based on the mass function
       parametrisation of \citet[][equation (\ref{eq:massfunction_fit})]{Tinker10}.
      The model, fitted over the mass range M123 (that is, excluding the low mass sample M0, see
      Table \ref{table:mass_ranges}) is in reasonable agreement
      with the measurements. Including the sample M0 (haloes with less than $80$ particles)
      to the fitting range leads to poor fits of the model. The fits at both redshifts differ by less than
      $5\%$ for $\ln(\nu) \lesssim 3$, confirming low redshift dependence from the measurements.
      The redshift dependence is stronger when lower masses are included in the fitting range, possibly because
      of redshift dependent noise in the low mass FoF detection.
      At larger masses (ln($\nu) > 3$) we find up to $10\%$ deviations, which are 
      comparable with the mass function errors.
      We have verified that our conclusions also hold for fits over the higher mass 
      range, M23, and different mass function binnings.
      A detailed analysis of the mass function fits, including fits of other mass function models over
      different mass ranges and different binnings as well as a comparison with fits compiled from the
      literature, can be found in the next section.
      
      \subsection{Mass function fits}\label{sec:mass_function_fits}
      
      In order to predict the halo bias from the mass function via the PBS approach we fit different
      mass function models to the measurements. Several systematic effects, such as the
      choice of the mass function model or the mass range
      over which the model is fitted can limit the accuracy of the PBS bias predictions
      \citep[e.g.][]{MS&S10, M&G11}. The objective of the subsequent analysis is to find out how strongly
      these effects impact the predicted linear, quadratic and
      third-order bias coefficients. In particular
      we aim to verify if the disagreement between PBS predictions for the linear bias and 
      the corresponding measurements from two-point correlations, presented in
      Section \ref{sec:bxi-vs-b_pbs}, is driven by possible shortcomings of the mass function fits.
      We therefore study in this subsection the fitting performances of different mass function models.
       
      The latest model in our analysis with the highest number of free parameters
      is the expression given by \citet{Tinker10} (hereafter referred to as Tinker model).
      It can be written as
      \begin{equation}
              \nu f(\nu)= A[1+(b\nu)^{a}] {\nu}^{d}e^{-c\nu/2},
              \label{eq:massfunction_fit}
      \end{equation}
     where $A, a, b, c, d$ are the free parameters. We have redefined
     the parameters so that fixing certain parameters delivers
      expressions which correspond to the mass function models suggested by \citet{PS74}, \citet{Sheth99}
      and \citet{warren06} (hereafter referred to as PS, ST and Warren model respectively).
      The corresponding parameter constraints are summarised 
      in Table \ref{table:massfunction_constraints} together with the abbreviations 
      for the reference of each model. This unification of notation
      allows  a more direct comparison between models.
     In Table \ref{table:massfunction_constraints}  we also
      propose a constrain, which constitutes a new mass function fit.
      Its advantage is that it has as many free parameters as the Warren model, but matches the
      mass function better when we fit over the whole mass range, as we show later.
      In our analysis we will focus on the models of ST, Warren, Tinker and our proposal.
       \begin{table}
        \centering
        \begin{tabular}{l | l l}
               model & reference & constraints \\
               \hline \\
               Tinker & \citet{Tinker10} & $A,a,b,c,d$ free \\
               Warren & \citet{warren06} & $d=0$ \\
               ST & \citet{Sheth99} & $c=b$, $d=1/2$ \\
               PS & \citet{PS74} & $a=0$, $c=1$, $d=1/2$ \\
               proposal & this work &$c=1$\\
         \end{tabular}
         \caption{constraints of parameters in equation (\ref{eq:massfunction_fit}) corresponding to
         different mass function models.  We refer to the models in the text using the abbreviations
         given in the left column.}
         \label{table:massfunction_constraints}
      \end{table}
      We determine the best fitting parameters for each mass function model by minimising
      \begin{equation}
      \chi^2=\sum^{N_{bin}}_{ij} \Delta_i \hat{C}_{ij}^{-1}\Delta_j,
      \label{eq:chisq}
      \end{equation}
      with $\Delta_i\equiv (X^{fit}_i-X_i)/ \sigma_{X_i}$ and $X=\nu f(\nu)$. $\hat{C}_{ij} $ and $\sigma_{X_i}$
      are derived from our new JK estimator, introduced in Appendix \ref{app:covar}.
      For searching the best fitting parameters we implemented a Monte Carlo Markov Chain algorithm 
      to explore the parameter space.
       
\begin{figure*}
      	\centerline{\includegraphics[width=140mm,angle=270]{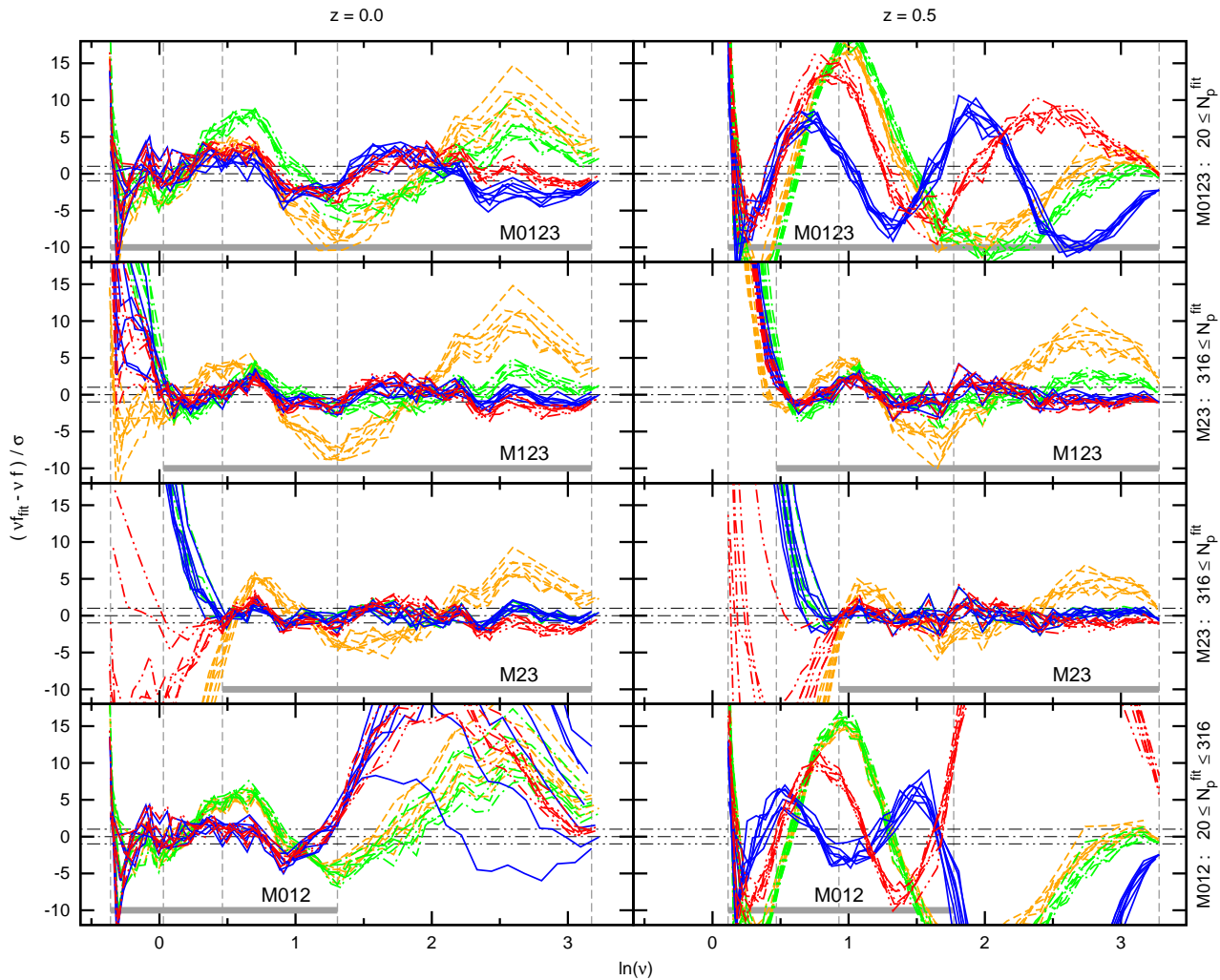}}
      	\caption{Significance of the deviations between mass function fits and measurements
      	versus the peak height $\nu \equiv \delta_{c}^2 /
        \sigma^2(m)$. Panels from top to bottom show results for
      	fits over the mass ranges M0123, M123, M23 and M012 respectively.
      	These ranges are marked by thick grey horizontal lines.
      	 Grey vertical lines denote the minimum and maximum peak heights of the different halo mass
      	 samples M0-M3. Dash-dotted horizontal lines denote $1 \sigma$ deviations between fits and measurements.
      	Results for the redshifts $z=0.0$ ($z=0.5$) are shown in the left 
      	(right) panels.
	Coloured lines show fits to the models of Tinker (solid blue), Warren (dashed-dotted green) and ST
	(dashed orange), while fits to our proposed model are shown as red dashed-double-dotted lines.
	For each model we show seven fits, which were derived from mass function measurements based on
      	dividing the whole mass range into $20,25,30,...,50$ bins.}
      	\label{fig:nufvsnu_mice_fits_signif}
      	\end{figure*}

      In Fig. \ref{fig:nufvsnu_mice_fits_signif} we show the significance of the deviations between
      the mass function measurements and the best fits by the different models.
      Results are shown at the redshifts $z=0.0$ and $z=0.5$, while at each redshift
      we fit the mass function over the different mass ranges, which are shown in Table \ref{table:mass_ranges}.
      The first includes all halo mass samples (M0123), the second
      and the third exclude the low mass samples (M123, M23) and the fourth mass
      range excludes the highest mass sample (M012).
      For each fitting range we show fits based on seven different mass function binnings, dividing the mass 
      range into $20,25,30,\ldots,50$ logarithmic bins.
      We find that the deviations between fit and measurement can vary with the binning. However,
      we also see trends which are independent of this systematic effect.
      
     All mass function models show a clear dependence of the best fit on the 
     chosen mass range, while this dependence is weaker for the ST model.
     %
      The Tinker parameterization is the
        model which best fits the measurements at both redshifts and
        all mass ranges.
      This can be attributed to the fact that it contains the highest number of free parameters.
      The best fit parameters for the Tinker model are given in Table \ref{table:tinker_fitparam}.
      For fits over the 
      whole mass range (M0123) our proposed mass function model seems to match the
      measurements almost as good as the Tinker model, while having one free parameter less.
      It also has the advantage of producing stable values for the parameters regardless
      of the range used for the fit.
      When the fits are performed only at the highest mass range (M23) 
      the Tinker and the Warren mass functions fit the data equally well, while 
      the proposed model is a slightly worse fit. The ST model delivers the poorest fits in all cases.
      At $z=0.5$ we find strong deviations between fits and measurements when 
      the fitting range includes the low mass sample M0. This indicates that the
      FoF detection of low mass haloes can be strongly affected by shot-noise,
      while this effect is stronger at higher redshift.
      
      \begin{table}
      \centering
        \begin{tabular}{c  c c}
         mass range   & halo masses [$10^{12}$$h^{-1}M_{\odot}$]   &  $N_p$\\
           \hline 
             M0123	&	$\ge 0.58$	&	$\ge 20$      \\
             M123	&	$\ge 2.32$	&	$\ge 80$      \\
             M23	&	$\ge 9.26$	&	$\ge 316$    \\
             M012	&	$0.58-100$	&	$20-3416$      \\
         \end{tabular}
          \caption{Halo mass ranges for mass function fits and clustering analysis. $N_p$ is the number of particles per halo.}
        \label{table:mass_ranges}
      \end{table}

     \begin{table}
        \centering
        \begin{tabular}{cccccccc}
      $z$ & mass range & $A$ & $a$ & $b$ & $c$ & $d$& $\frac{\chi_{min}^2}{d.o.f.}$ \\
      \hline
        $0.0$ & M0123 &  $0.28$ & $1.80$ & $0.22$ & $1.08$ & $0.47$ & $25.6$ \\
        $0.5$ & M0123 &  $0.31$ & $2.74$ & $0.20$ & $1.37$ & $0.87$ & $125.6$ \\        
        $0.0$ & M123   &  $0.24$ & $1.39$ & $0.22$ & $0.94$ & $0.34$ & $3.4$ \\
        $0.5$ & M123   &  $0.26$ & $1.70$ & $0.17$ & $0.98$ & $0.45$ & $3.5$ \\        
        $0.0$ & M23     &  $0.17$ & $1.10$ & $0.55$ & $0.85$ & $0.01$ & $1.5$ \\
        $0.5$ & M23     &  $0.22$ & $1.28$ & $0.34$ & $0.86$ & $0.05$ & $1.6$ \\
      \end{tabular}
         \caption{Best fit parameters for the Tinker mass function
           model taking covariance between different mass
         bins into account. We show fits over the mass ranges, M0123, M123 and 
         M23, defined in Table \ref{table:mass_ranges}, also 
        displayed in Fig. \ref{fig:Aabcd-chisqmin_numin_z0.0} for
        $z=0.0$.  We show the  mean of fits with different binnings.
         The corresponding standard deviations are typically at the $2\%$ level.}
        \label{table:tinker_fitparam}
     \end{table}
            
      	\begin{figure*}
      	\centerline{\includegraphics[width=85mm,angle=270]{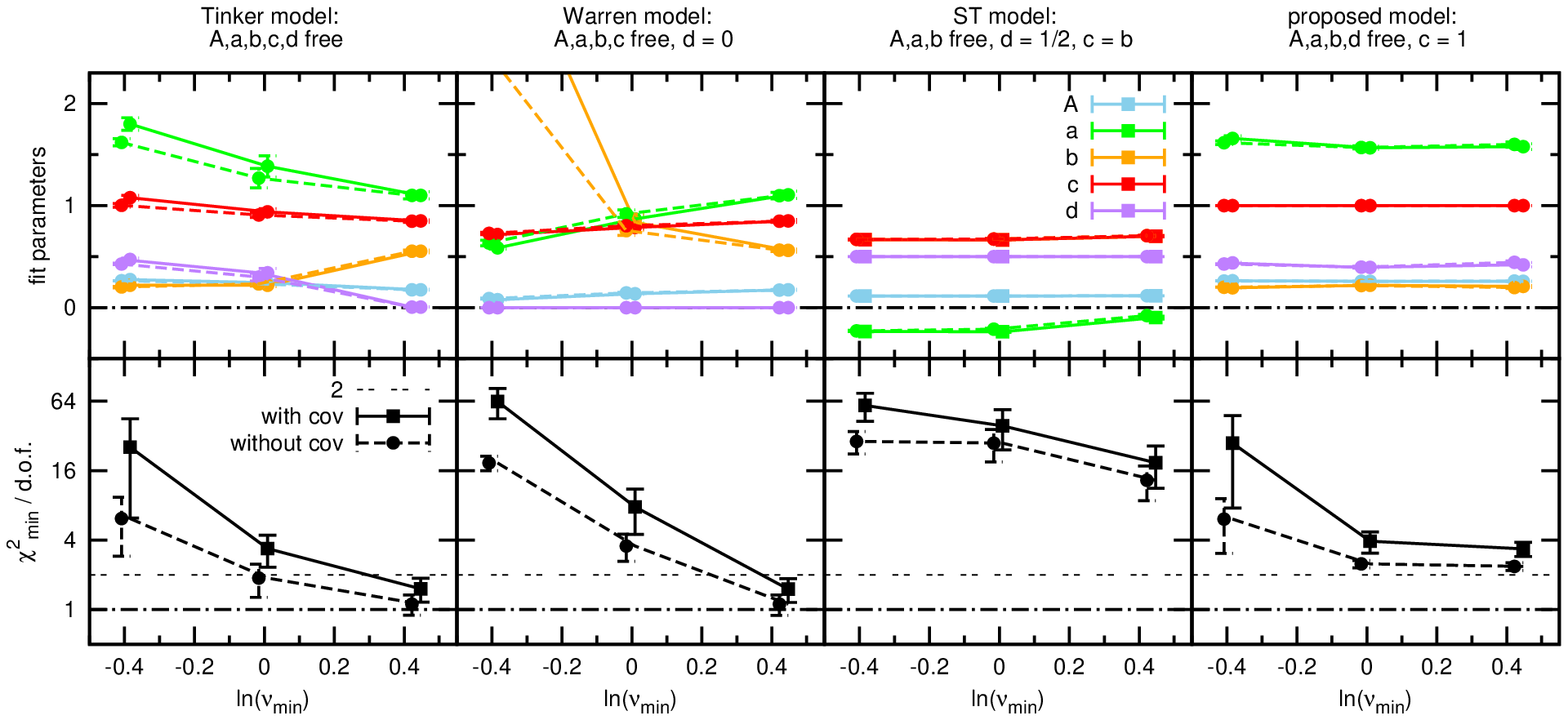}}
      	\caption{{\emph Top:} best fit parameters for different mass function models as a function
      	of the minimum peak height $\nu_{min}$ (corresponding to the mass ranges M0123, M123 and M23, defined in
      	Table \ref{table:mass_ranges}) used for fits at redshift $z=0.0$. Symbols show 
      	the means with standard deviations derived from seven mass function 
      	binnings (see Fig. \ref{fig:nufvsnu_mice_fits_signif}).
      Results from fits performed with and without taking the covariance
       between different mass function bins into account are connected with solid and 
      dashed lines respectively. In the latter case the symbols are slightly shifted to 
      the left for clarity.
      {\emph Bottom:} minimum $\chi^2/d.o.f.$ of the fits derived using our new 
      error estimator with $8^3$ JK samples. Note that errors can be smaller than the symbol size.
      We find similar results at redhsift $z=0.5$.}
      \label{fig:Aabcd-chisqmin_numin_z0.0}
      	\end{figure*}

      For studying the goodness of the best fits for the different mass function models we present their
      best fit parameters and the corresponding $\chi^2$ values per degree of freedom ($d.o.f.$)
      in Fig. \ref{fig:Aabcd-chisqmin_numin_z0.0}, where the $d.o.f.$ refer to the number of mass function
      bins used for the fit. Results are shown for fits over the mass ranges M0123,
      M123 and  M23, which correspond to the different minimum peak heights given by the x-axis.
      For clarity we show here only results at redshift $z=0.0$, while we find similar results at $z=0.5$.
      For each fit we show mean results with standard deviations from the seven mass binnings
      mentioned previously. In addition to the results derived by taking the 
      covariance between different mass function bins into account in the 
      fitting procedure we show results which were computed neglecting the 
      covariance. We find that neglecting the covariance can lead to different 
      best fit parameters, especially when the low mass range, where the off-diagonal elements of
      the covariance have the highest amplitudes, is included in the analysis (see Appendix \ref{app:covar}). 
      However, the bias predictions are only weakly affected by the negligence 
      of the covariance (see    Fig. \ref{fig:b1c2c3_pbs_allmodels}). The conclusions of this article
      about the comparison between bias predictions and measurements
      does not  dependent strongly on the covariance use.
      
      The $\chi^2/d.o.f.$ results, shown in the bottom panel of Fig. \ref{fig:Aabcd-chisqmin_numin_z0.0},
      are very high when the mass functions are fitted over the whole range (lowest minimum peak height).
      This poor performance, which is even apparent for the Tinker model with its 
      five parameters, is probably related to the fact that our mass function measurements are not 
      reliable in the low mass range. In fact the M0123 sample includes
      haloes with down to $20$ particles. For such low numbers of particles per 
      halo we expect strong systematic effect in the halo mass estimation and 
      therefore on the mass function \citep[e.g.][]{warren06,MoKraDaGott11}.
      Furthermore, the halo samples might be contaminated with spuriously linked FoF groups.
      If the analysis is performed using only the 
      high mass sample M23 (highest minimum peak height), the $\chi^2/d.o.f.$ values 
      for the best fit models drop down to values between unity and four. If we 
      perform the fits ignoring off-diagonal elements in the covariance matrix we obtain
      substantially lower $\chi^2/d.o.f.$ values, especially when the fits are performed over the whole mass 
      range. This demonstrates that the covariance cannot be neglected in the fit 
      for the evaluation of the fitting performance of a mass function model. This statement is
      even true in the high mass range where the covariance is
      dominated by shot-noise. This is important as the goodness of
      the fit is the way to validate the predictions.
      
      We also see in Fig. \ref{fig:Aabcd-chisqmin_numin_z0.0} that the $\chi^2/d.o.f.$ can change for different mass function binnings, 
      which can already be seen in Fig. \ref{fig:nufvsnu_mice_fits_signif}. This dependence 
      on the binning is also apparent when the off-diagonal elements of the covariance matrix are neglected
      in the fit. However, the best fit values of each model and the corresponding bias
      estimations are only weakly affected by this systematic effect.
      
      Interestingly the best fit parameters of the Tinker model have the same 
      values as the ones from the Warren model when the fit is performed on the 
      higher mass M23 sample. Consequently the minimum $\chi^2/d.o.f.$
      are the same in both cases. This indicates that the parameter $d$, which 
      is set to zero in the Warren model is not required for fitting the high mass 
      range, but becomes necessary, when the low mass range is included in the 
      fit.
     The $\chi^2/d.o.f.$ values of our proposed model are smaller than those for the Warren model
      for minimum peak heights of $\ln(\nu_{min}) \lesssim 0$ (M123). This agrees with the visual impression,
      gained from Fig. \ref{fig:nufvsnu_mice_fits_signif}, that our proposed model delivers better
      mass function fits than the model of Warren, unless the analysis is restricted to the highest
      mass range M23. We come to the same conclusion when analysing the mass function at $z=0.5$.

      \subsection{Mass function universality}\label{sec:mass_function_univ}

      In Fig. \ref{fig:massfctfit} we demonstrated that the mass function, when expressed in terms of
      the peak-high $\nu$, depends only weakly on redshift. To verify if this universality also holds for
      other cosmologies we compare our mass function fits to the Tinker and Warren 
      model with fits to the same models, compiled from 
       \citet[][Table 8]{warren06}, \citet[][Table 4, $\Delta_{200}$]{Tinker10},
       \citet[][Table 2]{crocce10} and \citet[][Table 2, FoF Uni.]{Watson13}. 
      \citet{crocce10} and \citet{Watson13} fit mass functions to the Warren 
      model. Note that \citet{crocce10} also used simulations from the MICE simulation suite,
      with the same cosmology as MICE-GC, but rely on the nested boxes approach to cover a
      similar mass range, while having a higher resolution in the low mass end than MICE-GC.
      \citet{Tinker10} used spherical overdensities to define haloes.
      A universal behaviour would not only be useful for PBS bias predictions, 
      but also for constraining $\sigma_8$ with galaxy luminosity functions,
      statistics of the initial density field and various other application 
      \citep[see e.g.][]{White02}.
      
\begin{figure}
      	\centerline{\includegraphics[width=100mm,angle=270]{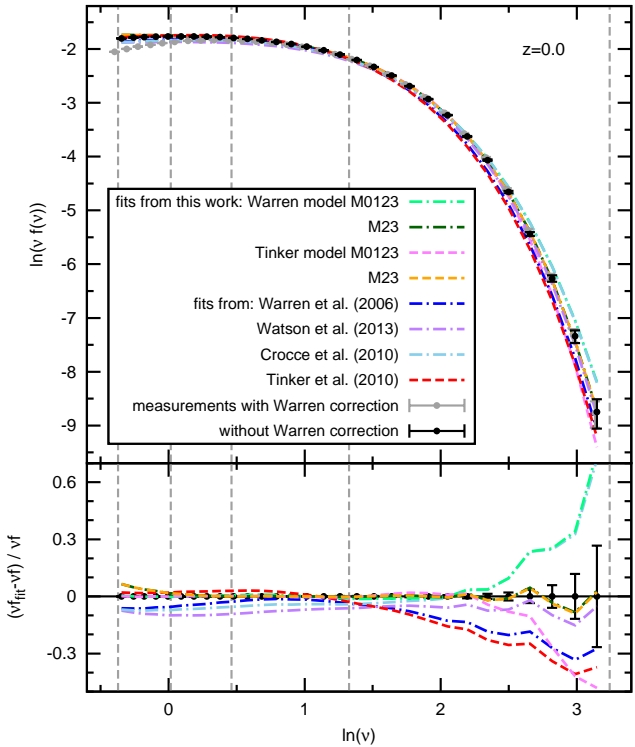}}
      	\caption{{\emph Top:} mass function fits compiled from the literature compared with
      	MICE-GC measurements and fits from this work over the whole mass range (M0123)
      	and the high mass range (M23) at $z=0.0$. Grey and black symbols show
      	measurements computed with and without Warren correction for halo masses respectively.
      All fits from this work are based on the latter.
      	{\emph Bottom:} relative deviations between fits and measurements in the same colour coding as the
      	top panel.}
      	\label{fig:massfct_comp}
      	\end{figure}

      We compare our mass function fits with those from the literature
      in Fig. \ref{fig:massfct_comp}. We find that the different mass 
      function fits agree at the $10\%$ level in the low mass end, but differ by 
      up to $60\%$ at high masses with a significance of about $2\sigma$ in terms of
      error in the measurement. Departures from
      universality are expected for different cosmologies but can also result from systematic effects,
      such as the halo mass definition \citep[e.g.][]{LaceyCole94, Sheth01, Jenkins01, White02, Reed07,
      Lukic07, Tinker08, crocce10, Courtin11, MoKraDaGott11, Bhattacharya11, Castorina14}. Furthermore,
      the fitting procedure affects the presented comparison as well.

      The comparison between the Warren fit from \citet{crocce10} and from 
      this work reveals the strong impact of the latter systematic effects on the fit.
      These two fits agree well in the high mass 
      end, when the fit is performed over the whole mass range M0123. 
      Interestingly we find that these fits differ more strongly from the 
      measurements in the high mass end than fits from other simulations.
      Excluding lower masses from the fit (M23) leads to a better agreement between our fit and 
      the measurement in the high mass end and therefore to a stronger difference 
      between the results from \citet{crocce10} and ours. The lower amplitude
      of the \citet{crocce10} fit at low masses indicates that the low halo mass 
      MICE-GC halo samples include more spuriously linked FoF groups, which
      can be expected from the low resolution as we concluded before in this 
      section. Furthermore, a lower mass resolution leads to an overestimation
      of halo masses. Correcting this effect as suggested by \citet{warren06}
      and done by \citet{crocce10} results in a decrease of the amplitude, which is shown
      as grey symbols in Fig. \ref{fig:massfct_comp}. The fact that our Warren corrected
      mass function is lower than all mass function fits in the low mass 
      range ($\ln(\nu) \lesssim 0$) indicates that the Warren correction leads to an 
      underestimation of halo masses when it is applied on FoF groups with order of $10$ 
      particles. For intermediate masses ($0 \lesssim \ln(\nu) \lesssim 2$) our Warren corrected measurements are in 
      better agreement with the results from \citet{crocce10} than those without 
      Warren correction. A comparison between the Warren corrected MICE-GC mass function at $z=0.0$
      and the \citet{crocce10} fit, presented by \citet{mice2}, also shows 
      higher amplitude of the prediction compared to the measurement in the 
      highest mass bin at $6$ $10^{14} M_{\odot}h^{-1}$ and an opposite trend for lower 
      masses.

%% file: sections/pbs_bias.tex
\section{PBS bias predictions}\label{sec:pbs_bias}

      The bias parameters $b_N$, introduced in equation (\ref{eq:biasfunction}), can be obtained from
      derivatives of the halo mass function via the PBS
      approach \citep{bardeen1986, cole1989, mo96}. Following \citet{Scoccimarro01} we derive
      the first-, second- and third-order bias parameters from the mass function fits as
      \begin{equation}
      b_1(\nu) = 1 + \epsilon_1 + E_1,
      \label{eq:massfunction_b1}
      \end{equation}
      \begin{equation}
      b_2(\nu) = 2(1+a_2)(\epsilon_1  +E_1) + \epsilon_2 + E_2,
      \label{eq:massfunction_b2}
      \end{equation}
      \begin{equation}
      b_3(\nu) = 6(a_2+a_3)(\epsilon_1+E_1) + 3(1+2a_2)(\epsilon_2 + E_2) \\
        + \epsilon_3 + E_3,
      \label{eq:massfunction_b3}
      \end{equation}
      where the parameters $a_2=-17/21$ and $a_3=341/567$ are given by the spherical collapse
      model. $E_1,E_2, E_3,\epsilon_1, \epsilon_2$ and $\epsilon_3$ are computed from
      the fitted parameters in the mass function models as shown in Table \ref{table:E123esp123}.
      Note that the non-linear bias parameters
      (equations (\ref{eq:massfunction_b2}) and (\ref{eq:massfunction_b3})), derived from the expressions in Table 
      \ref{table:E123esp123}, are here presented for the first time for the Tinker model.
      Applying the parameter constraints from Table \ref{table:massfunction_constraints} 
      delivers the equivalent expression for the PS, ST and the Warren models,
      as well as for our proposed model.
                  
       \begin{table}
        \centering
        \begin{tabular}{c|c}
         \hline \\
              $\epsilon_1 \equiv $ &
              $\frac{c \nu-2d}{\delta_c}$  \\ \\
              $\epsilon_2 \equiv $ &
              $\frac{c \nu(c \nu-4d-1 )+2d(2d-1)}{\delta_c^2}$  \\ \\
              $\epsilon_3 \equiv $ &
              $\frac{c\nu[(c\nu)^2 -6(d+1/2)c\nu  + 12d^2] - 8d^3 + 12d^2 -4d}{\delta_c^3}$ \\ \\
              $E_1 \equiv $ &
              $\frac{-2a}{\delta_c[(b\nu)^{-a} + 1]}$  \\ \\
              $E_2/E_1 \equiv $ &
              $\frac{-2a + 2c\nu -4d+1}{\delta_c}$ \\ \\
              $E_3/E_1 \equiv $ &
              $\frac{4a^2+12a(d-1/2)+2(2d-1)^2+4d(d-1)-6c\nu(2d+a)+3(c\nu)^2}{\delta_c^2}$
              \\ \\
         \end{tabular}
         \caption{Coefficients for computing halo bias parameters from the \citet{Tinker10} mass 
         function model via equations (\ref{eq:massfunction_b1}), (\ref{eq:massfunction_b2}) 
         and (\ref{eq:massfunction_b3}). $a, b, c$ and $d$ are the free parameters in the Tinker model.
         Bias predictions for other mass function models can be obtained by using the constraints
         for the fitting parameters, given in Table \ref{table:massfunction_constraints}.}
         \label{table:E123esp123}
      \end{table}
           
\begin{figure}
	\centerline{\includegraphics[width=200mm,angle=270]{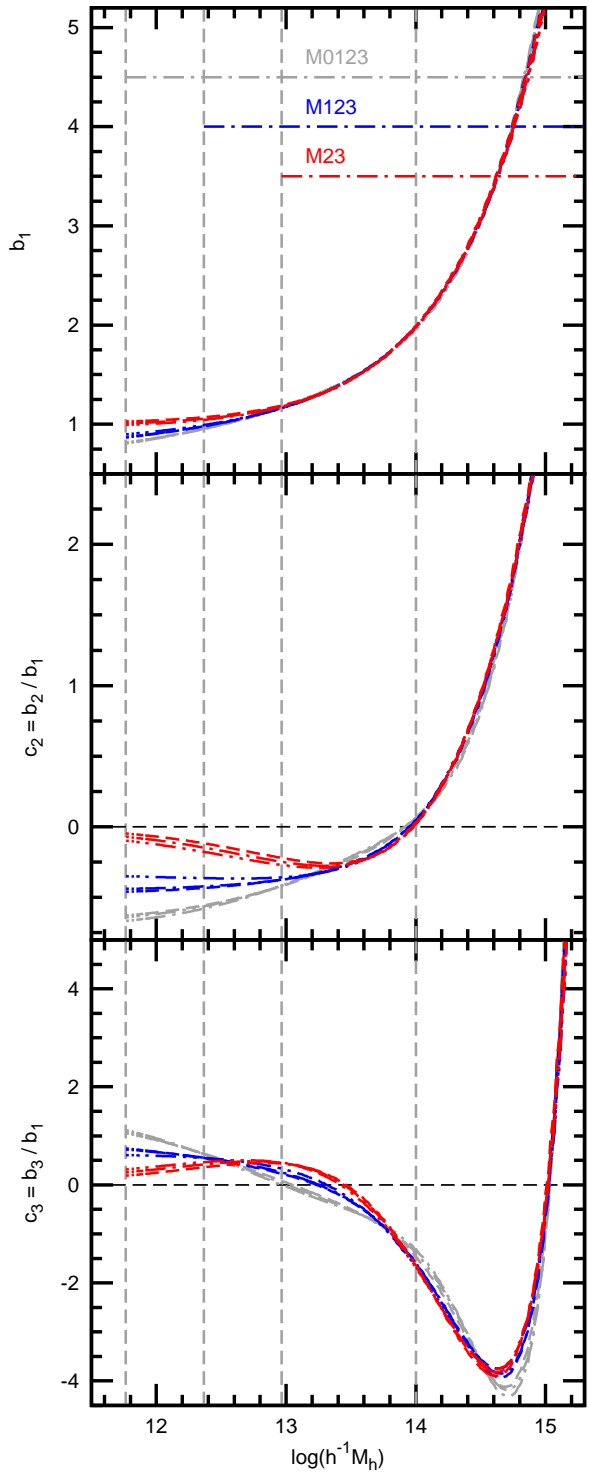}}
	\caption{Bias parameters $b_1$, $c_2 \equiv b_2 / b_1$ and $c_3 \equiv b_3 / b_1$
	(top, central and bottom panels respectively),
	derived from mass function fits of the Tinker model via the PBS approach at $z=0.0$.
	Grey lines show results based of mass function fits over the whole mass range {M0123}, blue and red lines show
	results from mass function fits which exclude the lowest and the two lowest mass samples
	(M123 and M23 respectively). Results based on fits to mass function measurements with
	$20$, $30$ and $40$ bins are shown as dashed, dashed-dotted and dashed-double-dotted
	lines respectively. Results derived from fits of other mass function models performed in this work
	are shown in Fig. \ref{fig:b1c2c3_pbs_allmodels}}
	\label{fig:b1c2c3_pbs}
	\end{figure}

      Predictions for $b_1$, $c_2 \equiv b_2 / b_1$ and $c_3 \equiv b_3 / b_1$,
      derived from the Tinker mass function fits at 
      $z=0.0$, are shown as a function of FoF halo mass in Fig. \ref{fig:b1c2c3_pbs}. 
      The results are based on mass function fits over the whole mass range (M0123)
      and fits over the higher mass ranges (M123 and M23). The $b_1$ predictions 
      for the different fitting ranges agree in the high mass end where the fitting ranges 
      overlap and the mass function fits agree as well (see Fig. \ref{fig:nufvsnu_mice_fits_signif}).
      In the low mass end we find a clear, but relatively weak dependence of the linear bias prediction on the fitting
      range. In the case of $c_2$ and $c_3$ this dependence is stronger and 
      reaches to higher halo masses. This indicates that second- and third-order derivatives of 
      the mass function, used to derive $c_2$ and $c_3$, cannot be measured as reliable as first-order
      derivatives, used to derive $b_1$.
      We see the same trends when employing the ST and Warren mass function models as well as for
      our proposed model, while in these cases the dependence on the fitting range is weaker
      (see Appendix \ref{app:pbs_bias}).
      We also find a similar behaviour of the bias predictions at $z=0.5$.
      %

\begin{figure*}
	\centerline{\includegraphics[width=125mm,angle=270]{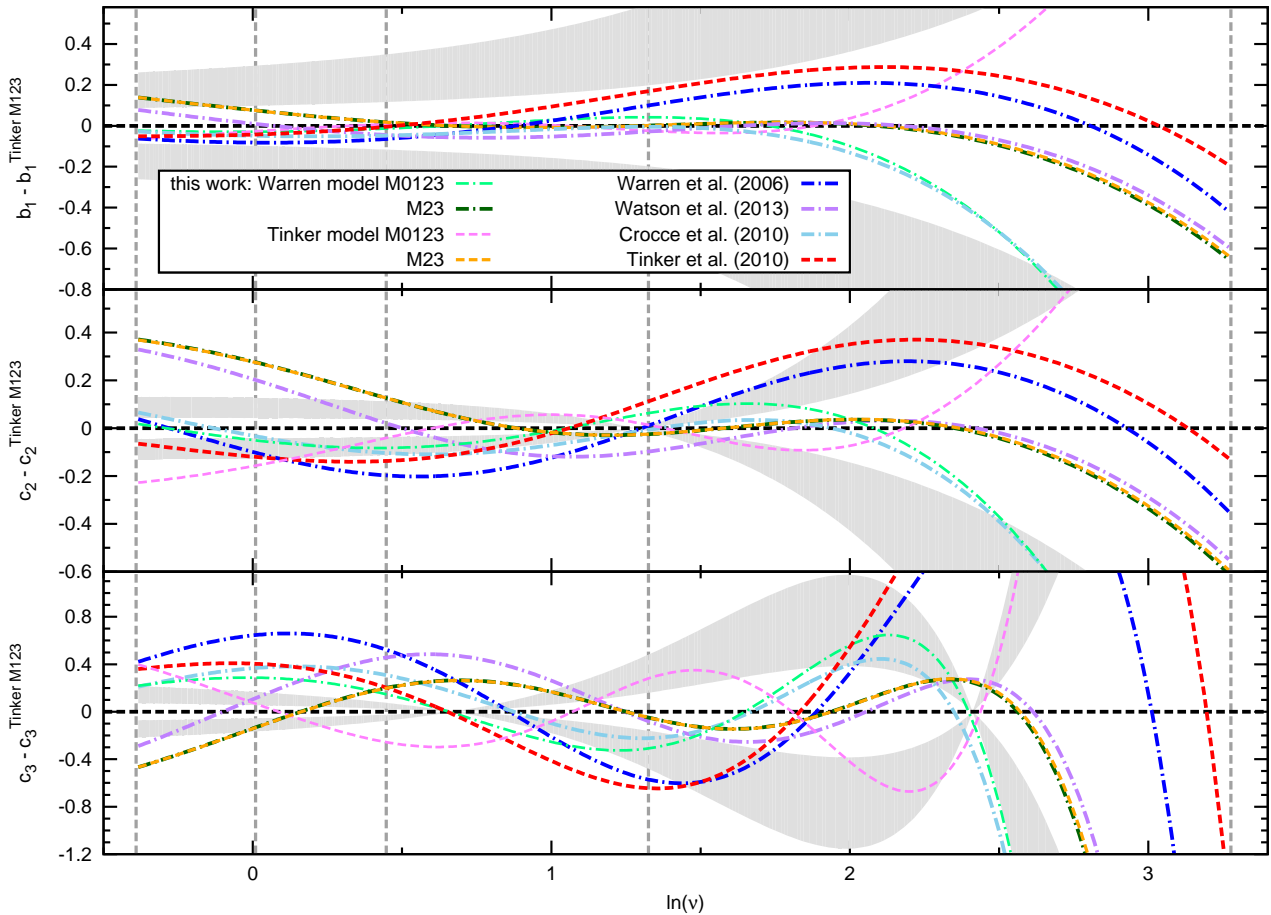}}
	\caption{Deviations of PBS predictions for $b_1$, $c_2$ and $c_3$ at $z=0.0$
	(top, central and bottom panel respectively) derived from various mass 
	functions fits with respect to the bias from our Tinker mass function fit over the range M123
	(shown as blue line in Fig. \ref{fig:b1c2c3_pbs}) as function of the peak height $\nu \equiv \delta_{c}^2 / \sigma^2(m)$.
	Deviations between $10-30 \%$ are marked by grey areas.
	The line color coding is the same as in Fig. \ref{fig:massfct_comp}. Vertical dashed lines
	denote the $\nu$ limits of the halo mass samples M0-M3.}
	\label{fig:deltab1c2c3_pbs}
	\end{figure*}

      The absolute deviations between bias prediction from the Tinker mass function, fitted over the range
      M123 and other predictions are shown in Fig. \ref{fig:deltab1c2c3_pbs}. These 
      other predictions are based on Tinker and Warren fits over different mass 
      ranges and fits for the same models compiled from the literature. We do not 
      show relative deviations to avoid singularities at the zero crossings of $c_2$ and $c_3$.
      For the linear bias we find absolute deviations between the different predictions of
      $\Delta b_1 \simeq 0.2$, which roughly corresponds to relative deviations of around 
      $10\%$. The relative deviations for $c_2$ and $c_3$ are around $50\%$, but can go
      up to more than $100\%$.
     Mass function fits over the high mass range M23 to the Tinker and Warren models 
     deliver almost identical bias predictions, which can be expected since also 
     the fitted parameters are very similar (see Fig. \ref{fig:Aabcd-chisqmin_numin_z0.0}).
     In the high mass end these two bias predictions agree with prediction from the fit to the Warren mass function
     given by \citet{Watson13}.
     Comparing our results to those of \citet{crocce10} we find a reasonable agreement
     for bias predictions based on the Warren model fitted over the whole mass range M0123.

      \subsection{Universal relation between bias parameters}\label{sec:b1c2c3relation}

      A universal behaviour of the mass function, as studied in Section \ref{sec:mass_function_fits},
      would suggest that the bias parameters, derived from the mass function are universal as well,
      when they are expressed as a function of peak height $\nu$.
      Our comparison with the literature shows that both, the mass function 
      from different simulations and the bias parameters derived from these mass functions 
      (especially $c_2$ and $c_3$)
      can differ significantly from each other. These disagreements might not only arise 
      from different cosmologies, but also systematic effects, as discussed previously.
      
      \begin{figure}
      \centerline{\includegraphics[width=140mm,angle=270]{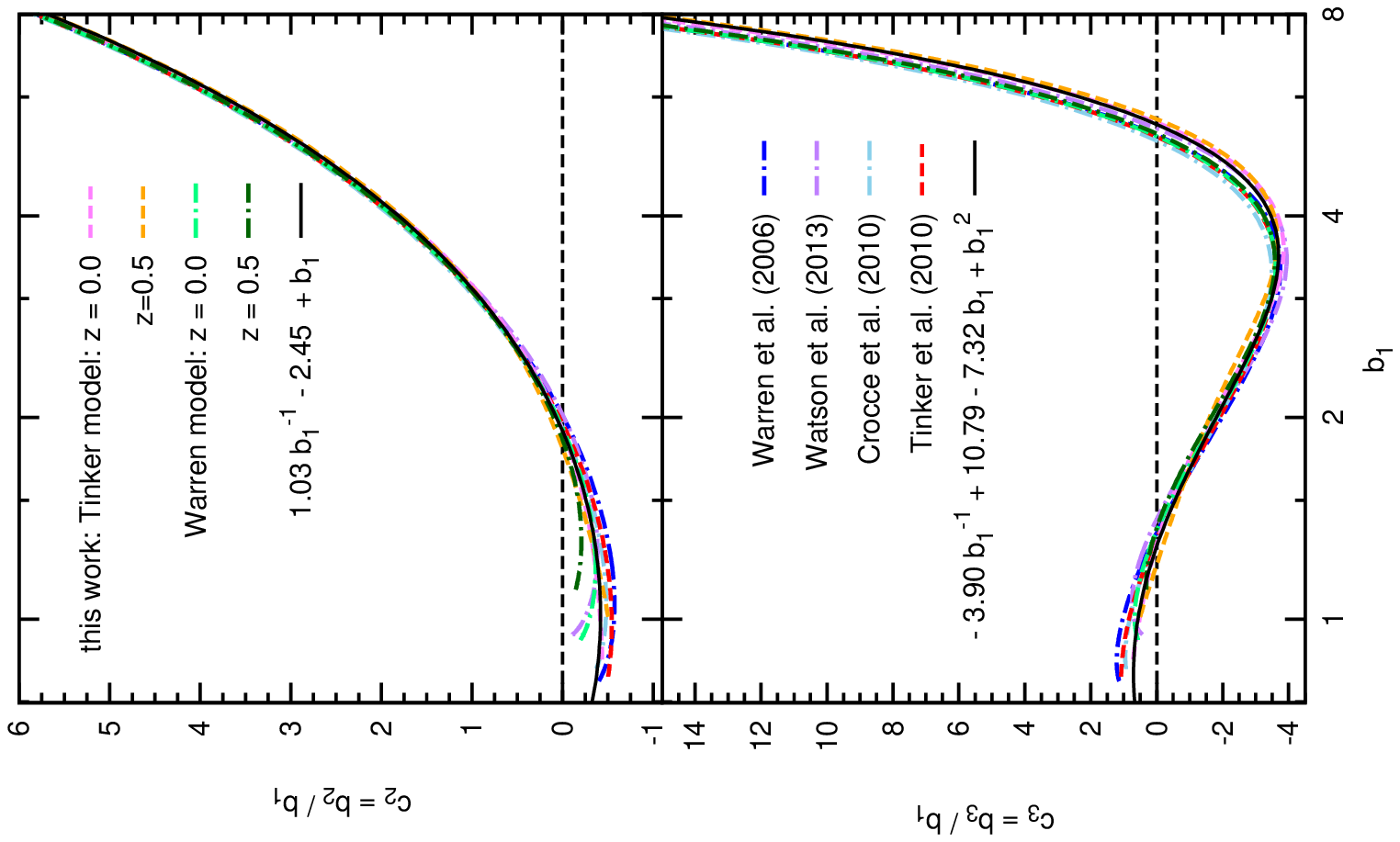}}
      	\caption{Second- and third-order bias parameters, $c_2 \equiv b_2/b_1$ and $c_3 \equiv b_3/b_1$, as a 
      	function of the linear bias parameter $b_1$, predicted from the PBS model (top and bottom panel
      	respectively). Results from this work are based on mass function fits 
      	over the mass range M123. Results from the literature are shown in the same colour coding
      	as in Fig. \ref{fig:massfct_comp}. Note that the mass function
      	fits from \citet{crocce10} and \citet{Watson13} are based on the Warren model. Black 
      	solid lines show polynomials (equation (\ref{eq:b1b2_b1b3_relation})), which were fitted to the PBS predictions of the Tinker 
      	model, based on MICE-GC mass function fits from this work at $z=0.0$ (magenta dashed line)
      	with rms per degree of freedom of $0.02$ and $0.12$ and for $c_2$ and $c_3$ respectively.}
      	\label{fig:b1-c2-c3}
      	\end{figure}
	
      We now aim to verify the universality of the relation between the bias parameters.
      Such a universal behaviour would be useful for reducing uncertainties in linear 
      bias measurements from third-order statistics \citep[e.g.][]{M&G11,paper1}.
      In Fig. \ref{fig:b1-c2-c3} we show the PBS
      prediction of the second- and third-order bias parameters, $c_2$ and $c_3$, as a function of
      the prediction for the linear bias $b_1$. We find a $10\%$ agreement for the $b_1-c_2,c_3$
      relations for large values of the linear bias ($b_1 \gtrsim 1.5$).
      These relations appear to be well described by second- and third-order 
      polynomials in the case of $c_2 \equiv b_2/b_1$ and $c_3 \equiv b_3/b_1$ respectively, 
      with
     \begin{equation}
            b_N = \sum_{n=0}^N \alpha_n b_1^n,
           \label{eq:b1b2_b1b3_relation}
      \end{equation}
      as we demonstrate in the same figure. This finding can be expected from expressing
      $b_2$ and $b_3$ as functions of $b_1$ with the PS model (Table \ref{table:massfunction_constraints}). 
      For this model the parameters $\alpha_n$ can be directly predicted as
      $(\alpha_0, \alpha_1, \alpha_2) = (0.51, -2.21, 1)$ for $ b_2=b_1 c_2$ and 
      $(\alpha_0, \alpha_1, \alpha_2, \alpha_3) = (-1.49, 8.02, -6.64, 1)$ for $ b_3=b_1 c_3$. 
      However, we find smaller rms values with respect to the Tinker and Warren predictions for
      the $b_1-c_2$ and $b_1-c_3$ relations, when we leave $\alpha_{n < N}$ as free parameters.
      We show values for $\alpha_n$ from fits to the Tinker predictions in Fig.
      \ref{fig:b1-c2-c3}.
      
      For predictions based on our fits over the whole mass range, M0123, we find deviation from 
      this universal behaviour, while these results involve the low mass samples
      which we found to be unreliable previously, possibly due to low mass resolution and
      noise in the halo detection. For lower $b_1$ values the different predictions differ
      more strongly from each other. However, a weakly universal relation, especially between $b_1$ 
      and $c_2$, might already help to improve $b_1$ constraints from third-order 
      statistics as these two parameters are usually treated as independent.
      A comparison of the $b_1- b_2$ relation, predicted by the PBS with measurements from combined
      second- and third-order clustering was presented by \citet{Saito2014}, who also find that this relation
      is consistent with redshift independence. We will pursue the study of this relation
      with different measurements of $b_1$ and $c_2$ in a future analysis
      (Bel, Hoffmann \& Gazta\~naga in preparation).

%% file: sections/bpbs_vs_bxi.tex
\section{Bias prediction versus measurements from clustering}\label{sec:bxi-vs-b_pbs}

      In the previous section we found that the PBS bias predictions depend on 
      the employed mass function model and the mass range over which the 
      models are fitted. We now aim to
      verify how the predictions for the linear bias, $b_1$, in these different cases compare to linear
      bias measurements from the two-point halo-matter cross-correlation.
      A comparison of second- and third-order
      bias parameter predictions with other measurements will be presented in a future analysis
      (Bel, Hoffmann \& Gazta\~naga in preparation).

\begin{figure*}
         \centering
         \includegraphics[width=100mm, angle = 270]{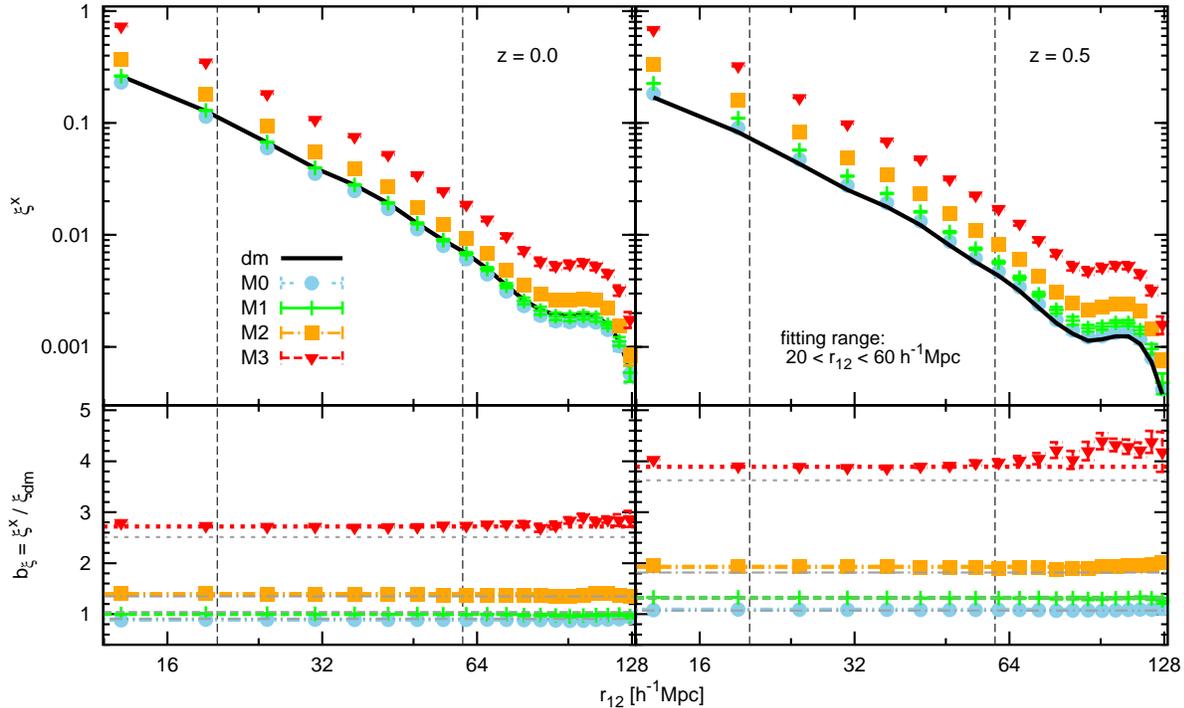}
         \caption{\emph{Top}: two-point correlation $\xi$ of the MICE-GC
           dark matter field (continuous lines) and the two-point halo-matter cross-correlation
           for the halo mass samples M0-M3 (blue circles, green crosses, orange squares and red
           triangles respectively) in the comoving outputs at redshift $z=0.0$ (left) and $z=0.5$ (right)
           as a function of scale $r_{12}$.
        \emph{Bottom}: linear bias parameter $b_{\xi}$ derived from the two-point correlations via
        equation (\ref{eq:bxi}). Coloured lines are $\chi^2$-fits between $20-60$ $h^{-1}$Mpc.
         The minimum $\chi^2$ values per degree of freedom are
         $1.05, 1.96, 1.54, 0.23$ for M0, M1, M2, M3 respectively at $z=0.0$ and
         $0.79, 1.46, 1.23, 0.77$ for M0, M1, M2, M3 respectively at $z=0.5$.
         Grey lines show PBS bias predictions.
         The same figure for the auto-correlation is shown in \citet{paper1}.}
         \label{fig:b1_2pccfits}
      \end{figure*}

      The two-point cross-correlation between halo- and matter density fields, 
      $\xi^{\times}$, can be measured as the mean product of smoothed fluctuations
      $\delta(\bf r) \equiv (\rho(\bf r)- \bar{\rho})/\bar{\rho}$ of each
      density field, $\rho(\bf r)$, at the positions $\bf r_1$ and $\bf  r_2$ as a function of the 
      scale $r_{12} \equiv |\bf r_1- \bf r_2|$,
      
      \begin{equation}
		\xi^{\times}(r_{12}) \equiv \langle  \delta_{h}(\bf r_1) \delta_{m}(\bf r_2) \rangle.
		\label{eq:def_2pcc}
	\end{equation}
      The measurements for the four halo mass samples M0-M3 at the redshifts $z=0.0$ and $z=0.5$
      are shown in the top panels of Fig. \ref{fig:b1_2pccfits}.
      The amplitude increases with halo mass as expected from the PBS 
      predictions. The growth of matter fluctuations further contributes to a
      change with redshift. At around $110$ $h^{-1}$Mpc $\xi^{\times}$ shows a local
      maximum which results from baryonic acoustic oscillations in the initial
      power spectrum of the simulation.
      
      A relation between the two-point halo-matter cross-correlation and the 
       two-point matter auto-correlation,
       $\xi_m(r_{12}) \equiv \langle  \delta_{m}(\bf r_1) \delta_{m}(\bf r_2) \rangle$,
       via the halo bias can be obtained
      by inserting the local bias model from equation (\ref{eq:biasfunction}) into equation 
      (\ref{eq:def_2pcc}),
      	\begin{equation}
      \xi^{\times}(r_{12})  \simeq ~b_1~\xi_{m}(r_{12}) + o[\xi_{m}].
      \end{equation}
      At large scales ($r_{12} \gtrsim 20 \ h^{-1}$Mpc) we expect the higher-order contributions $o[\xi_{m}]$ to be 
      negligible, which allows for measurements of the linear bias as
	\begin{equation}
      b_{\xi}(r_{12}) \equiv \frac{\xi^{\times}(r_{12})}{\xi_{m}(r_{12})}
      \simeq b_1.
      \label{eq:bxi}
      \end{equation}
      The measurements of $b_\xi$ are shown in the bottom panel of Fig. \ref{fig:b1_2pccfits}. 
      We fit $b_{\xi}$ between $20-60$ $h^{-1}$Mpc, where 
      the scale-independence is a good approximation. Non-linear terms impact $b_{\xi}$ at smaller scales,
      but also around the scale of baryonic acoustic oscillations. Comparing these bias 
      measurements from the cross-correlation to those from the auto-correlation, 
      shown in \citet{paper1}, we find that non-linearities
      have a stronger effect  on the autocorrelation.
      However, differences in the bias from auto- and cross-correlations
      are small compared to differences between these measurements and the PBS predictions.
      We present a detailed analysis on the impact of non-linearities on bias from second-order statistics
      in \citet{paper2}.

      To compare the PBS predictions for the linear bias with the
      measurements from the two-point correlation we calculate the average
      bias prediction in each of the mass samples M0-M3, weighted with the halo number density 
      $n(m)$,
	\begin{equation}
	b(M)= \frac{\int_{M_{low}}^{M_{up}} b(m) n(m) dm}{\int_{M_{low}}^{M_{up}}  n(m) dm}.
	\label{b_pbs_mean}
	\end{equation}
	$M_{low}$ and $M_{up}$ are the lower and upper limits of each mass sample M,
	given in Table \ref{table:halo_samples}.
	PBS $b_1$ predictions, based on fits to the Tinker model over the mass range M123,
	are compared with the $b_{\xi}(r_{12})$ measurements in the bottom panel of Fig. \ref{fig:b1_2pccfits}.
	For the high mass samples M2 and M3 we find clear deviations between
	measurements and predictions as the latter are significantly
        too low, on all scales.
		       
\begin{figure*}
      	\centerline{\includegraphics[width=150mm,angle=270]{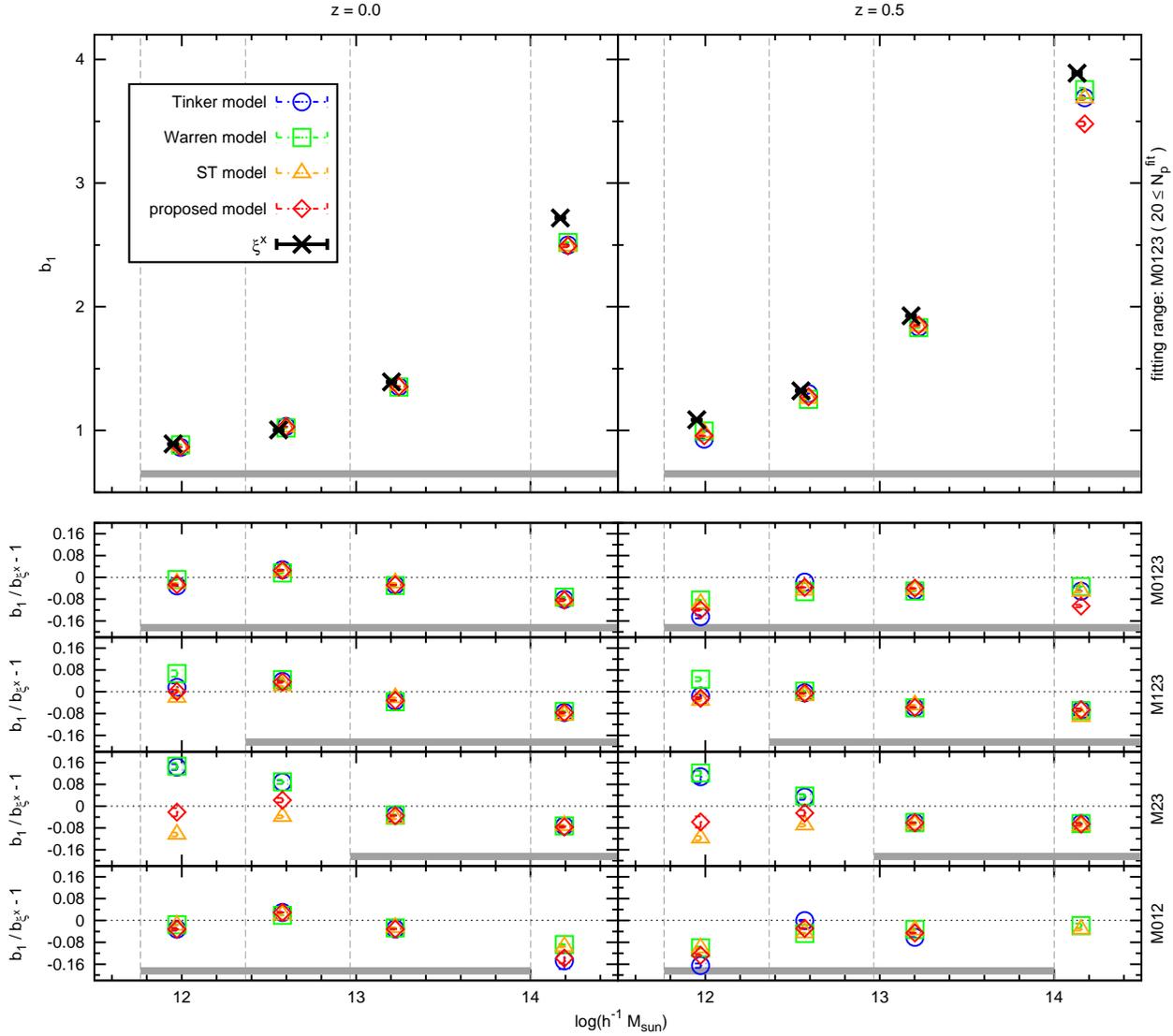}}
        \caption{
        \small {\it Top}: linear bias parameters $b_1$ for the halo mass samples 
        M0-M3 in the MICE-GC comoving outputs at redshift $z=0.0$ (left) and $z=0.5$ (right)
        versus the mean halo mass of each sample. Measurements from the
        two-point halo-matter cross-correlations, $\xi^\times$, via equation 
        (\ref{eq:bxi}), $b_{\xi^\times}$,
        are shown as black crosses with $1\sigma$ errors. PBS predictions, derived from MICE-GC
        halo mass function fits from this work are shown as coloured symbols,
        which are slightly shifted to the left on the mass axis for clarity. The mass range over 
        which the mass function is fitted is marked by a thick grey horizontal line.
        Error bars for the predictions are standard deviations derived from the seven different
        mass functions binnings shown in Fig. \ref{fig:nufvsnu_mice_fits_signif}.
        \small {\it Bottom}:  relative difference between $b_\xi$ and the PBS predictions.
        The different panels show results for predictions based on mass function fits over the mass ranges
        M0123, M123, M23 and M012, from the top to the bottom respectively.}
      	\label{fig:b1_pbs-xi_mass}
      	\end{figure*}

	The dependence of these deviations on the mass function model and the mass
	range in which the models are fitted is shown in Fig. \ref{fig:b1_pbs-xi_mass}.
	Fitting the mass function over the whole mass range, M0123, delivers $b_1$ predictions 
	which tend to be $1-15\%$, below the measurements, except for the low mass samples
	M0 and M1 at $z=0.0$, for which we find up to $5\%$ deviations in the opposite 
	direction. 
	We find the strongest variations between bias predictions from different models
	when, i) the low mass range at $z=0.5$ is included in the mass function fitting
	range, or ii) when the bias is predicted for mass samples which are not within the 
	fitting range (e.g. bias predictions for the mass sample M1, based on fits over the mass range M23).
	The first case i) might be explained by noise, contaminating 
	the FoF halo detection, which results in the poor mass function fits 
	shown in Fig. \ref{fig:nufvsnu_mice_fits_signif} (see discussion in Section \ref{sec:mass_function_fits}).
	In this latter figure we also see that the mass function fits outside the fitting range can strongly differ
	for different models. This could cause the strong differences in the bias predictions,
	described above as case ii).
	We do not see that the deviations between PBS bias predictions and
	$b_{\xi}$ measurements decrease when the analysis is
	restricted to the higher mass range. This is true for both redshifts ($z=0.0$ and $z=0.5$)
	and consistent with results from \citet{M&G11}.
	
	However, restricting the fitting range to the higher mass range M23 we find a good 
	agreement between the linear bias predictions from different mass functions 
	models at $z=0.0$ and $z=0.5$.
	The fact that the fitting performance strongly differs for the different 
	models (see Fig. \ref{fig:Aabcd-chisqmin_numin_z0.0}), while all models predict a linear
	bias with similar deviations to the measurements, suggests that the goodness of
	the mass function fit is not the only reason for these deviations, as mentioned 
	in the introduction to this article.
	These results line up with reports of \citet{MS&S10}, who also find the linear PBS bias
	prediction to lie below measurements from the power spectrum and two-point 
	correlations, especially at high halo masses. As in our case their result is independent
	of the employed mass function model and the way it is fitted to the measurements.
	
	Furthermore, these authors investigate if differences between the predictions and
	measurements are related to the mass definition of haloes. They therefore 
	perform their analysis using FoF groups with different linking lengths, as well as
	spherical overdensities to define halo masses. In both cases they find that the PBS model 
	underpredicts the linear bias measurements. In fact one could expect that the 
	halo mass should be higher than those of FoF groups in order to match
	the PBS predictions (since shifting the $b_\xi$ measurements in Fig. \ref{fig:b1_pbs-xi_mass}
	to higher masses would decrease the deviations between measurements and predictions).
	However, halo masses defined by spherical overdensities tend to be below those 
	of FoF groups \citep{Tinker08}. This should lead to higher measurements of the 
	linear bias for spherical overdensities within a given mass range than 
	corresponding measurements for FoF groups, as found by \citet[][]{Tinker10}.
	The $10\%$ underprediction of linear bias measurements by the PBS model, which
	we see for high mass haloes in Fig. \ref{fig:b1_pbs-xi_mass}, is therefore probably a 
	lower bound. The consideration above also suggests that applying the Warren correction on the FoF 
	masses could increase the differences between the PBS bias predictions and 
	the measurements. Hence, these differences might not only be related to the halo mass 
	definition, but also to assumptions of the PBS model, such as spherical collapse or a local 
	bias relation \citep[e.g.][]{Schmidt13, Paranjape13}.
	The conclusion, that bias predictions are only weakly dependent on the employed
	mass function model does not hold for the higher-order bias predictions $c_2$ and $c_3$
	(see Fig. \ref{fig:b1c2c3_pbs}, \ref{fig:deltab1c2c3_pbs} and \ref{fig:b1c2c3_pbs_allmodels}).

	\subsection{Bias ratios}

      The degeneracy between the growth of matter fluctuations and the bias of observed
      galaxy samples is one of the largest uncertainties for constraints of cosmological models 
      derived from large-scale structure observations. With estimations of the typical host
      halo masses of such galaxy samples the PBS model can be employed to predict the bias 
      of these samples to break the growth-bias degeneracy. Besides the galaxies host halo mass
      estimation, the inaccuracy of the PBS bias prediction constitutes an additional source of error in 
      this approach. Here we aim to quantify the impact of such inaccuracies on measurement
      of the linear growth factor. The considered growth measurements are based on the ratio of the
      correlation functions of galaxy samples at two different redshifts, $z_1$ and $z_2$, 
      multiplied with the inverse ratio of the bias of these samples \citep[see e.g.][]{paper1}.
      The bias ratio needs to be estimated or predicted, while its uncertainties 
      propagate linearly into the growth measurements.
      
\begin{figure}
      	\centerline{\includegraphics[width=105mm,angle=270]{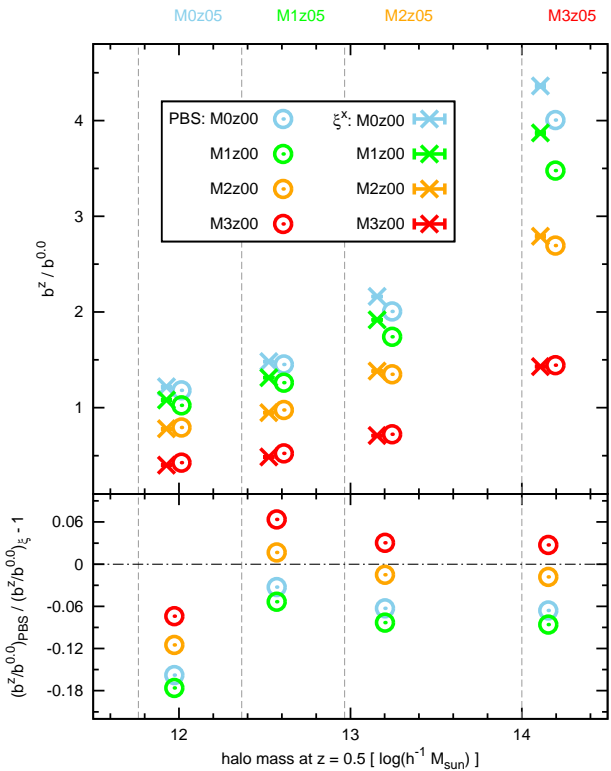}}
        \caption{\small {\it Top}: The ratio of halo bias at the redshifts
        $z=0.0$ and $z=0.5$, which could be used to measure the linear growth factor of matter 
        fluctuations. The PBS predictions, shown as open circles, are based on fits of the Tinker model
        to mass function measurements over the mass range M123. Measurements from the two-point
        correlation are shown as crosses with $1\sigma$ errors, derived from propagating the error
        of $b_{\xi}$ at the two redshifts. Note that errors are smaller than the symbol size.
        \small {\it Bottom}: relative deviations between predictions and measurements.}
      	\label{fig:b1_pbs_ratios}
      	\end{figure}

      In Fig. \ref{fig:b1_pbs_ratios} we show the PBS bias ratio predictions for the
      redshifts $z_1=0.0$ and $z_2=0.5$ and all combinations of the four halo mass 
      samples M0-M1 at these two redshifts.
      The predictions are based on fits of the Tinker model  to the mass function of the mass range M123,
      which we found to be reliable at both redshifts previously. We find an 
      overall variation of $5-10 \%$ for the higher mass range M123, while deviations are stronger
      when the low mass sample M0 at redshifts $z=0.0$ is taken into account. This 
      variation is stronger than uncertainties expected from the $b_{\xi}$ measurements.
      The strong deviations for the low mass range are expected due to the poor mass 
      function fit including M0 at $z=0.5$ (see Section \ref{sec:mass_function_fits}).
      The error in the bias ratio will propagate into $5-10 \%$ 
      error of the growth factor measurement.
     This uncertainty is lower than the 
      uncertainties found for growth measurements based on bias ratio estimations
      from the three-point correlation \citep[see ][]{paper1}.
      However, the estimation of the galaxies host-halo mass will introduce additional
      limitation in breaking the growth-bias degeneracy. Furthermore, the precision of any HOD
      fitting or mass interpretation from clustering measurements will be affected at similar level.

%% file: sections/conclusion.tex
\section{Summary and Conclusion}\label{sec:conclusion}

We investigated bias predictions from the PBS model, derived via fits to MICE-GC FoF mass functions. 
The accuracy of this model was tested by comparing its predictions for the linear bias to
direct measurements from two-point correlations. In order to verify how the bias 
predictions are affected by the goodness of the mass function fit, we study the performance of
four mass function models, fitted over different mass ranges at the redshifts $z=0.0$ 
and $z=0.5$. These fits are based on a new mass function error and covariance estimator.

We show that the 
models of \citet{PS74}, \citet{Sheth99} and \citet{warren06} are special cases of the mass function
expression suggested by \citet{Tinker10}, as they correspond to certain
values of free parameters in the Tinker model (see Table \ref{table:massfunction_constraints}).
This finding motivated us to propose a new model by fixing a different free parameter.
The fitting performance of the mass function models, quantified by the minimum $\chi^2$ 
values per number of analysed mass function bins ($d.o.f.$), shows strong variations among
different models and fitting ranges (see Fig. \ref{fig:Aabcd-chisqmin_numin_z0.0}).
All models match the measurements better when the low mass range is excluded from
the analysis. This indicates resolution effects, given that we analyse
FoF groups with down to $20$ particles.
We find that the model of \citet{Tinker10} shows the best overall performance, which can be
expected since it contains the highest number of free parameters.
Our proposed model delivers results similar to those from the Tinker model
when the whole mass range is analysed, while it has one free parameter less. These
two models outperform the model of \citet{warren06} for fits over the whole mass range. 
A restriction to the high mass range ($\ge 2.32$ $10^{12} h^{-1}M_{\odot}$) leads to very similar
fitting performance of the \citet{warren06} and \citet{Tinker10} models with minimum $\chi^2/d.o.f.$
values close to unity, while our proposal is slightly worse. Fits to the model of \citet{Sheth99} show the
most significant deviations to the measurements in all cases. These findings are independent 
of the mass function binning.
We find that the inclusion of the covariance into the analysis substantially
increases the minimum $\chi^2$ values of the best fits and also has an impact on
the best fit parameters. However, our PBS bias predictions are only very weakly 
affected by the mass function covariance, especially when the higher mass range is 
analysed, where errors are shot-noise dominated.

The results described above can be affected by the way the mass function errors and the
covariance between different mass function bins are estimated. We therefore conducted a detailed
study of these quantities which is presented in Appendix \ref{app:covar}.
Given the one MICE-GC realisation, we rely on the internal JK error estimator which we compared to
theory predictions. The comparison reveals
that the JK method is in good agreement with the predicted mass function error only in the
shot-noise dominated high mass range ($\gtrsim 5 \ 10^{14} M_{\odot}$), but
overestimates the predictions by up to $80\%$ in the lower mass range, where the errors are
dominated by sampling variance. We show that this difference arises
because the standard JK estimator assumes a wrong scaling
relation between sampling variance and sample volume.
By introducing an improved scaling relation, predicted from the linear matter power 
spectrum, we are able to propose a new mass function error estimator.
Deviations between errors of our new estimator and the predictions are less than 
$10\%$ (see Fig. \ref{fig:nofm_errcomp}). The advantage of the new estimator with respect to predictions is that it does
not rely on a model for halo bias and does not depend on the power spectrum
normalisation. This approach to JK error estimations can also be applied to
other statistics, such as two-point correlation functions (Hoffmann et al. in preparation).

The presence of non-zero off-diagonal elements in the mass function covariance suggests
that a similar covariance can be found in the luminosity function or the stellar
mass function, as reported in the literature \citep[e.g.][]{Smith12, Benson2014}.
The latter work demonstrated that the negligence of covariance in the stellar
mass function significantly affects parameter constraints in semi-analytic models
of galaxy formation. In this case a correct estimation of the error and covariance
might be important for a correct interpretation of observations within such models.

Our FoF mass function measurements show no significant ($\lesssim 5\%$)
change between the redshifts $z=0.0$ and $z=0.5$ for haloes with
more than $80$ particles (corresponding to a lower halo mass limit
of $2.32$ $10^{12} h^{-1}M_{\odot}$, see Fig. \ref{fig:massfctfit}). When including lower 
masses, the redshift dependence is stronger, possibly because of redshift 
dependent noise in the low mass FoF detection. In order to investigate a dependence on 
cosmology we compare our results with mass function fits from the literature.
We find variations between $10 \%$ in the low mass end and up to $40-60\%$ in the high mass end
(see Fig. \ref{fig:massfct_comp}).
This finding is in agreement with other studies on departures from the mass function universality
(see Section \ref{sec:mass_function_univ} for references). The advantage of the MICE-GC
simulation is the large $\sim(3 \ \text{Gpc}/h)^3$ volume, which leads to smaller errors in the high mass
range and therefore allows for a more significant assessment of the aforementioned variations in the mass function amplitude.

Our analysis demonstrated that numerical and systematic effects, such as halo mass definition, resolution effects
or the fitting procedure can contribute to variations in the mass function amplitude with a similar impact as differences
in the cosmology (see Section \ref{sec:mass_function_fits}). This result indicates that understanding
such numerical and systematic effects is important for discriminating different cosmologies
using the mass function. In fact it was shown in the literature that uncertainties in the mass
function in the order of magnitude that we find in our analysis, can strongly affect constraints
on the matter density, the dark-energy equation of state, the power spectrum amplitude $\sigma_8$
or neutrino properties from surveys such as DES or Euclid \citep[see e.g.][]{crocce10,Wu2010,
Costanzi13, Weinberg2013, Appleby2013, Basse2014, Bocquet2015}.

After comparing fits from different mass function models to MICE-GC measurements,
we study the bias prediction, derived from the best performing model
of \citet{Tinker10}. Note that the non-linear bias parameter expressions for this model 
are presented in this work for the first time.
We find that the bias prediction depends on the mass range over which the model was fitted
as the amplitude of the linear bias predictions varies by around $10\%$ for different fitting ranges.
For the second- and third-order bias parameters the amplitude can vary by more than $50\%$. 
These dependences of the bias predictions on the fitting range are comparable with
variations obtained when employing fits to other mass function models.
Furthermore we find deviations with similar amplitudes in a comparison with bias prediction
from mass function fits to other simulations, compiled from the literature
(see Fig. \ref{fig:deltab1c2c3_pbs}).

A universal behaviour of the mass function would suggest that the
bias parameters, derived from the mass function are universal as well.
Despite the strong variation among different bias predictions we find a tight universal
relation between $b_1$ and $c_2$ or $c_3$ for $b_1 \gtrsim 1.5$ across different
simulations and mass function models.
For smaller $b_1$ values, these relations are more dependent on the mass function fit, but still quite tight.
Using the PS mass function model we derive
that the second- and third-order bias parameters $b_2$ and $b_3$ can be 
expressed as second- and third-order polynomials of the linear bias $b_1$
(see Fig. \ref{fig:b1-c2-c3}).
These findings suggests that the linear bias can, at least, constrain the non-linear bias parameters.
This could be used to improve the linear bias measurements from third-order statistics.

A common application of the PBS model is to predict the linear bias from 
clustering. We measured the latter directly from the two-point halo-matter cross-correlation
at large scales in the MICE-GC and compare it to the PBS predictions. The 
comparison was conducted using four different mass samples at the redshifts
$z=0.0$ and $z=0.5$. Excluding the low mass sample M1 with less than $80$ particles per halo
from the analysis, which we expect to be affected by noise, we find that the linear bias, predicted
from the PBS, model lies $5-10\%$ below results from the two-point correlation
(see Fig. \ref{fig:b1_pbs-xi_mass}).
This effect is similar at the redshifts $z=0.0$ and $z=0.5$ and 
independent of the employed mass function model and the way it is
fitted to the measurements, confirming previous findings \citep{MS&S10, M&G11}. 
Including the low mass sample delivers similar results,
but with a larger scatter among the models. From the analysis in the higher mass
ranges we conclude that shortcomings in the fitting performance of the mass function model
are not the main reason for the discrepancy between PBS predictions for the linear bias and
the corresponding measurements from clustering. An alternative reason for such discrepancies
might be given by the overestimation of halo masses by the FoF algorithm, as those tend to be larger 
than halo masses of spherical overdensities \citep{Tinker08}. However, from our 
results in Fig. \ref{fig:b1_pbs-xi_mass} we conclude that shifting the linear bias 
measurements of FoF halo samples to lower masses would increase
the deviations between measurements and PBS predictions. Hence, if FoF masses are 
overestimations of the halo masses described by the PBS model then the 
differences between linear bias predictions and measurements, found in this 
analysis, constitute lower bounds for the inaccuracy of the predictions. 
This indicates that simple assumptions of the PBS model, such as a local bias model or
spherical collapse might limit the accuracy of the linear bias predictions. We 
will present a comparison between the non-linear bias parameters from 
predictions and measurements in Bel, Hoffmann \& Gazta\~naga (in preparation).

The $5-10\%$ deviations between linear bias predictions and measurements will affect at
similar level the precision of any HOD fitting or mass interpretation from clustering measurements.
We demonstrate the impact of these deviations on growth measurements from two-point correlations.
Such measurements are based on the ratio of the linear bias at two different redshifts. 
Ignoring the unreliable low mass range we find $5-10\%$ deviations between PBS 
predictions for the bias ratios and measurements from the two-point correlation. 
This inaccuracy would propagate linear into measurements of the linear growth factor, based on
PBS bias predictions.

%% file: sections/appendix.tex
\section{Mass function measurements}\label{app:mass_function_measurement}
      The mass function measurements are based on a rewritten form of equation \ref{eq:massfunction_def}:
      \begin{equation}
      \nu f(\nu) \equiv   \frac{\langle m \rangle}{\bar \rho}\frac{dn(m,z)}{dlg \ m} \frac{dlg \ m}{dln \ 
      \nu},
      \label{eq:massfunction_measurement}
      \end{equation}
      where $\langle m \rangle$ is the mean halo mass in each logarithmic mass bin.
      If the mass bins are chosen to be exactly equal in logarithmic space, the
      mass function amplitude slightly oscillates in the low mass range due to mass resolution 
      effects. Since the errors are smallest in the low mass range this artefact can significantly
      affect the fits, causing a strong dependence of the fits on the number of
      mass function bins. Aiming to minimise this mass discreteness effect we 
      determine the minimum and maximum number of particles per halo in each logarithmic mass 
      bin, $(N_p^{max}$ and $N_p^{min}$ respectively. The width of the mass bin in then
      recalculated as $m_p(N_p^{max}-N_p^{min}+1)$, where $m_p$ is the particle mass.
      The value of $\nu$ of each bin is calculated from the mean mass of 
      haloes in the bin. The term $\frac{dlg \ m}{dln \ \nu}$ in equation \ref{eq:massfunction_measurement} is derived directly from
      equation(\ref{eq:sigma}).

      \section{Covariance}\label{app:covar}
      
      In order to fit the mass function we estimate the errors and the covariance between different mass bins.
      A direct measurement of these quantities would require a large set of 
      independent realisations of the simulation. Since just one realisation of the MICE-GC simulation
      was run we estimate the errors and the covariance using the Jack-Knife (hereafter referred to as JK)
      sampling technique. To validate these estimations we compare the results to 
      theoretical predictions, which we will describe first.
      
      \subsection{Covariance prediction}
      
     Following \citet{crocce10} we derive the covariance prediction for the comoving halo number
     density from the linear bias relation at large scales,
      \begin{equation}
       n(m,{\bf r}) = \bar{n}[1 + b_1(m) \delta_m({\bf r})] + \delta n^{sn}(m,{\bf r}),
      \label{eq:lin_bias_model}
      \end{equation}
      where $n(m,{\bf r})=N(m,{\bf r})/V_{tot}$ is the number density of haloes with mass $m$ in a 
      volume (in our case the simulation volume) around position ${\bf r}$, $\delta_m({\bf r})$ is the
      matter density contrast in the same volume and
     $b_1=\delta_h/\delta_m$ is the linear halo bias factor
     (as before $m$ refers to the matter density field when it appears as lower index
     and to the halo mass when it is used as variable).
     The last term, $\delta n^{sn}(m,{\bf r})$, corresponds to noise.
      We will assume $\delta n^{sn}$ to be Poisson shot-noise and therefore independent of ${\bf r}$.
      The predictions for the unconditional mass function can be  related to those for the halo number
      density via equation (\ref{eq:massfunction_def}). For the sake of simplicity the following 
      considerations are based on the latter.
      The covariance matrix for number densities of haloes in the mass bins $i$ and $j$
      is defined as
      \begin{equation}
       C_{ij} \equiv \langle \Delta_i \Delta_j\rangle = \frac{1}{N_{samp}}  \sum_k^{N_{samp}}  \Delta_i^k \Delta_j^k,
      \label{eq:def_covar}
      \end{equation}
      where $\Delta_i^k \equiv (n_i^k - \bar{n}_i)$  and $\langle \ldots\rangle$ denotes the average 
      over $N_{samp}$ statistically independent volumes $k$
      (note that the $\Delta$ introduced here is not related to the $\Delta$ used in equation (\ref{eq:chisq}) for
      calculating the $\chi^2$ values of mass function fits).
      Inserting the expression for the number density $n_i$ of haloes in mass bin $i$ from equation (\ref{eq:lin_bias_model}) leads to
      \begin{equation}
       C_{ij} =
       \bar{n}_i \bar{n}_j b_i b_j \langle\delta_m^2 \rangle +
       \langle \delta^{sn}_i \delta^{sn}_j \rangle.
      \label{eq:covar_prediction}
      \end{equation}
      The variance of matter fluctuations $\langle \delta_m^2 \rangle = \sigma_m^2(m_{tot})$ can be derived
      from the power spectrum via equation (\ref{eq:sigma}), while $m_{tot}$ is the total
      mass within the volume in which the mass function is measured (in our case the total mass in the
      simulation). Since this mass corresponds to a very large smoothing radius we can compute $\sigma_m^2$ 
      from the linear power spectrum. The sampling variance contribution to the covariance is 
      therefore given by
      \begin{equation}
       C^s_{ij} = \bar{n}_i \bar{n}_j b_i b_j \sigma_m^2(m_{tot}) 
      \label{eq:covar_prediction_sampling}
      \end{equation}
      If the noise term $\delta^{sn}$ is Poissonian it averages out when taking the mean over many
      independent volumes. The contribution of  shot-noise to the covariance is then given by
      \begin{equation}
       C^{sn}_{ij} = \delta_{ij} \frac{\sqrt{\bar{n}_i  \bar{n}_j}}{V_{tot}},
      \label{eq:covar_prediction_shot-noise}
      \end{equation}
      while here $\delta_{ij}$ is the Kronecker delta. Based on these considerations we can write the
      total covariance as
       \begin{equation}
       C_{ij} = C^s_{ij} + C^{sn}_{ij}.
      \label{eq:covar_tot}
      \end{equation} 
      A more formal derivation for this relation is given by \citet{Smith2011}, see also
      \citet{Robertson10, Valageas11, Smith12}.
      The diagonal elements of the covariance matrix correspond to the predictions for the
      mass function variance,
       \begin{equation}
       \sigma_i^2 = C_{ii}
      \label{eq:var-covar}
      \end{equation}
       as given by \cite{crocce10}.
         For fitting the mass function we work with the normalised covariance
         $\hat{C}_{ij}\equiv{C}_{ij}/(\sigma_i\sigma_j$) and differences normalised to $\sigma_i$.
         (note that here $\sigma_i$ refers to the variance of the mass function in the mass bin $i$
         and not to the variance of the matter field, $\sigma_m$).

      \subsection{Jack-Knife estimation of covariance}\label{sec:cov_prediction}

      For mass function fits in observations the covariance prediction is of limited use since
      it requires knowledge about the bias and the power spectrum in advance.
      This problem might be solved with an iterative approach for the fit, starting
      from an initial guess for the power spectrum and the linear bias factor. Another possibility to
      obtain the covariance without knowledge of the bias and the power spectrum is to estimate it
      with the JK sampling technique. Testing this approach we construct $N_{JK}$ JK samples by subtracting cubical
      sub-volumes (hereafter referred to as JK cells) with the size $V_{tot}/N_{JK}$ from the total
      simulation volume $V_{tot}$. The basic assumption of the JK approach is 
      that the error scales with the size of the subtracted volume \citep[e.g.][]{norberg09}.
      We follow the common approach by rescaling the covariance with the factor
      $(N_{JK}-1)$, which leads to
      
      \begin{equation}
       C^{JK}_{ij} \equiv (N_{JK}-1) \langle \Delta_i \Delta_j \rangle =
       \frac{N_{JK}-1}{N_{JK}}  \sum_k^{N_{JK}}  \Delta_i^k \Delta_j^k.
      \label{eq:def_JKcovar}
      \end{equation}
     Again $\Delta_i^k = (n_i^k - \bar{n}_i)$, but now $\langle \ldots\rangle$ is the average 
      over the different JK samples $k$,  $n_i^k$ is the comoving number density
      of haloes in the mass bin $i$ in each JK sample and $\bar{n}_i$ is 
      the corresponding halo number density in the whole simulation volume.
      Note that the rescaling factor, {\bf $(N_{JK}-1)$}, is only weakly justified and can be improved, as we show
      in Section \ref{sec:newJK}.
     As in the case of the predictions the diagonal elements of $C_{ij}^{JK}$ are the JK estimation
     for the variance ($\sigma_i^2 = C_{ii}$) and we normalise
     $\hat{C}^{JK}_{ij}\equiv C^{JK}_{ij}/(\sigma^{JK}_i\sigma^{JK}_j)$.
     Note that we can use the JK approach for studying the covariance between 
     low and high mass bins because of the large mass range of the MICE-GC 
     simulation. This analysis would not be possible using nested boxes, 
     where the different mass ranges are covered by different realisations with different box sizes
     \citep[e.g.][]{warren06, crocce10,Tinker10}.
     
    \subsection{Covariance prediction versus Jack-Knife estimation}

\begin{figure}
      	\centerline{\includegraphics[width=85mm,angle=270]{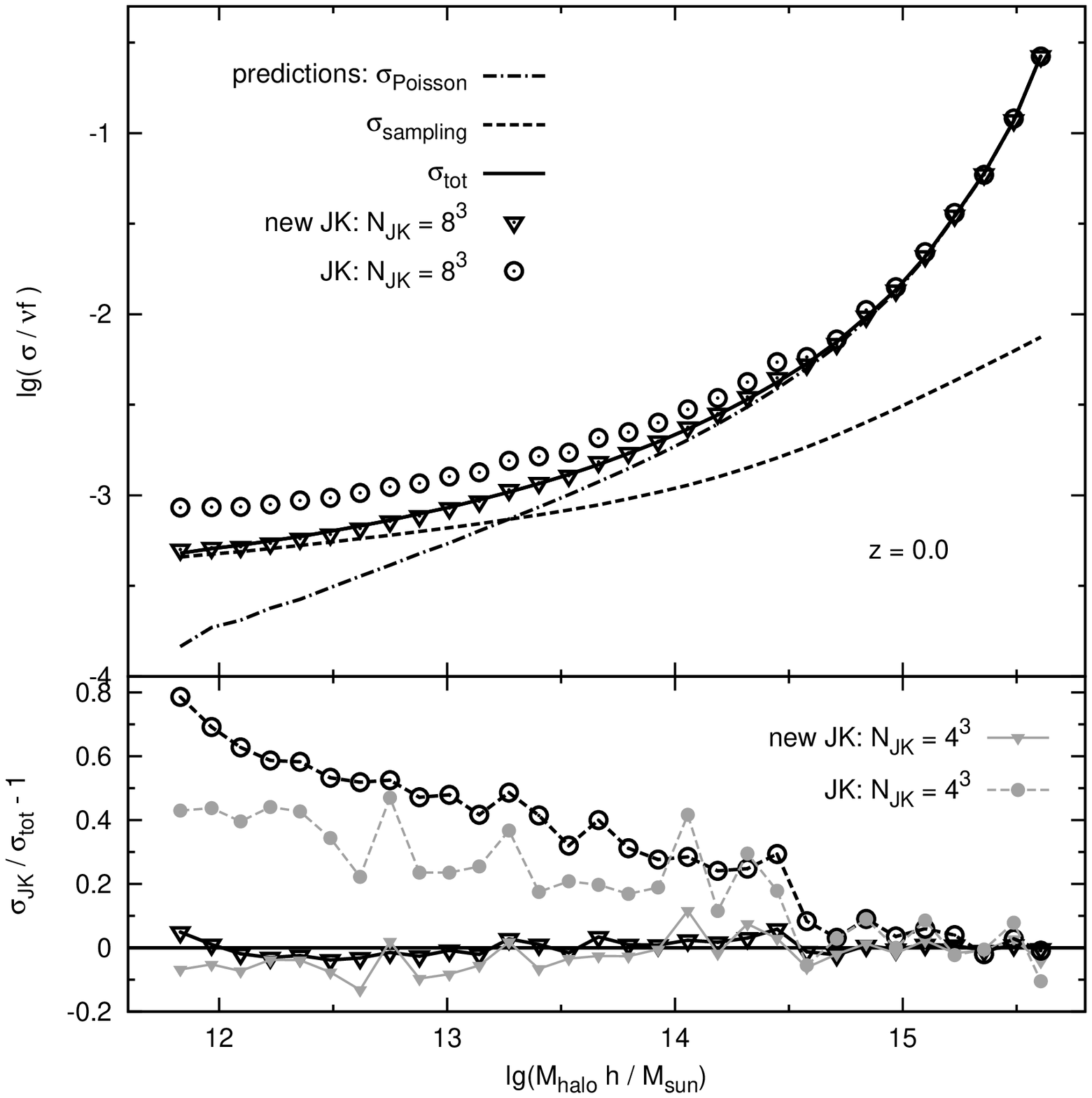}}
      	\caption{
	\emph{Top}: Relative errors in the mass function. Lines 
         show the  predictions, including the sampling variance (dashed line)
      	and the Poisson shot-noise contribution (dash-dotted) together
        with the resulting total error (continuous line),
      	derived from the equations (\ref{eq:covar_prediction_sampling}) - (\ref{eq:covar_tot})
      	(with $\sigma_i^2=C_{ii}$). Open circles show the standard
        JK error estimation (equation (\ref{eq:def_JKcovar})) and
      	open triangles show the errors from a new JK estimation (equation (\ref{eq:new_jkcovar})).
      	Both estimations are based on $8^3$ JK cells. \emph{Bottom}:
     	relative deviations between the JK and the total predicted error. 
        The symbol types corresponds to those in the top panel. Results
     	derived from $4^3$ cubical JK cells are shown as grey solid symbols.}
      	\label{fig:nofm_errcomp}
      	\end{figure}

    A comparison between the error prediction from equation (\ref{eq:covar_tot}) and the
    corresponding JK estimation from equation equation (\ref{eq:def_JKcovar}) (with $\sigma_i^2 = C_{ii}$)
     is shown in Fig. \ref{fig:nofm_errcomp} for the redshift $z=0.0$.
     The error predictions are based on linear bias predictions from mass 
     function fits to the Tinker model over the whole mass range,
     for which we expect uncertainties of around $10\%$ (see Section \ref{sec:pbs_bias}).
     From the prediction we expect the error to be dominated by sampling variance in the low mass end 
    and by shot-noise in the high mass end. At halo masses of $M_h\simeq 2 \ 10^{13}M_{\odot}$ 
    both sources are predicted to contribute equally to the total error.
    The JK error estimation is in good agreement with the predictions in the high mass 
    end ($M_h\gtrsim 5 \ 10^{14}M_{\odot}$). This indicates that the JK method is working well for different JK
    cell volumes when the error is dominated by shot-noise. Furthermore, the shot-noise is well
    described by a Poisson distribution.
    At halo masses lower than $5 \ 10^{14}M_{\odot}$
    we find the JK error to be up to $80\%$ higher than the prediction.
    
    This overestimation is consistent with results reported by \citet{crocce10} 
    using the same simulation box size as the MICE-GC, while for smaller simulation 
    boxes they find the JK error to be lower than the prediction.
    The fact that the overestimation of the JK error in the low mass end is larger for smaller JK 
    cells indicates that the JK assumption of a linear relation between errors and volume
    is inadequate when sampling variance is the dominating source for error.
    However, increasing the size of the JK cells results in a smaller 
    number of samples and therefore a stronger noise on the estimated error.
    In Fig. \ref{fig:nofm_errcomp} we also show a new JK error estimation, which 
    is in good agreement with the predictions at all masses. This new JK error is based on 
    an improved scaling between the sampling variance in a JK cell and in the whole 
    simulation box using the linear power spectrum, as explained in Section \ref{sec:newJK}.
    
\begin{figure}
      	\centerline{\includegraphics[width=190mm,angle=270]{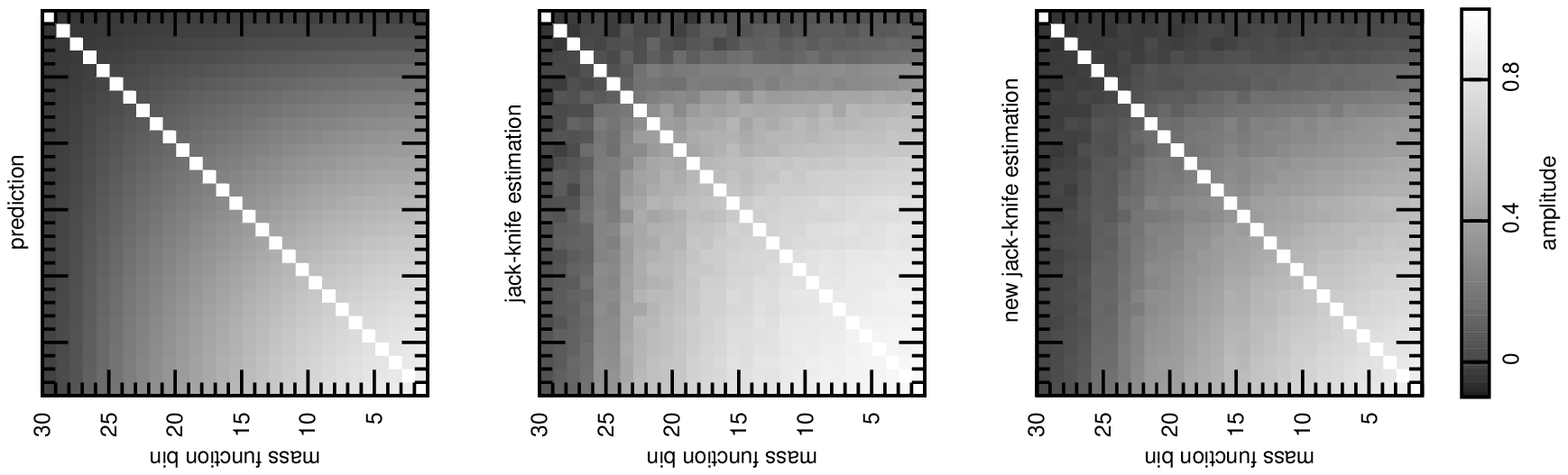}}
      	\caption{Normalised covariance between mass function bins at redshift $z=0.0$, 
      	derived from predictions (top, equations (\ref{eq:covar_prediction_sampling}) - (\ref{eq:covar_tot})),
      	the standard JK estimator (center, equation(\ref{eq:def_JKcovar})) and the new JK estimator
      	(bottom, equation (\ref{eq:new_jkcovar})). The estimations are based on $8^3$ JK cells.}
      	\label{fig:covar}
      	\end{figure}

      In Fig. \ref{fig:covar} we compare the normalised covariance of the mass function
      between the mass bins $i$ and $j$, predicted via
      equation (\ref{eq:covar_tot}) with the JK estimation from equation (\ref{eq:def_JKcovar})
      using $8^3$ JK samples at $z=0.0$. The shape of the covariance is in good
      agreement with results from \citet{Smith2011}.
      The low mass bins are highly correlated because of sampling variance, while high mass bins are 
      uncorrelated as their errors are dominated by shot-noise.
      For the comparison of the variances we find a reasonable agreement between
      the prediction and the JK estimations, especially in the high mass end. In 
      the low mass end the covariance seems to be overestimated by the JK 
      approach, while the new JK method reproduces the prediction well.
           
\begin{figure}
      	\centerline{\includegraphics[width=75mm,angle=270]{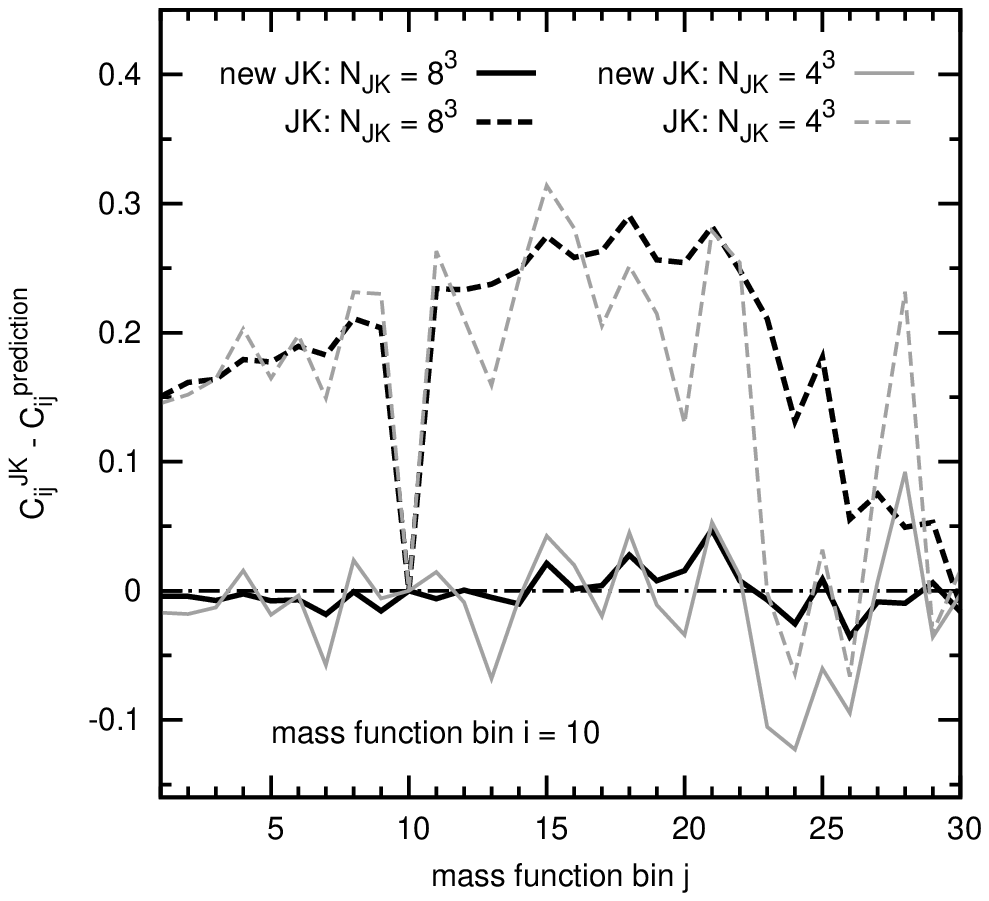}}
      	\caption{Deviations between the standard and the new JK covariance estimation
      	(solid and dashed lines respectively) and the prediction, fixing one bin to $i=10$
      	and varying the other. We find similar results for different values of $i$.
      Black and grey lines show results based on $8^3$ and $4^3$ JK cells respectively.}
      	\label{fig:covar_diff}
      	\end{figure}

      We show a more detailed comparison of the covariance amplitudes in Fig. \ref{fig:covar_diff}, 
      fixing one mass bin $i$ and varying the second mass bin $j$.
      For $8^3$ JK cells we find the normalised JK covariance amplitudes
      to be higher than the predictions with differences of up to $0.3$.
      Using larger JK cells this overestimation slightly decreases, while results become more noisy.
      Again the improved estimation is in better agreement with the prediction.
      We have verified that our conclusions also hold for redshift $z=0.5$.
      
    \subsection{Improved JK estimator}\label{sec:newJK}

    We now aim to understand the disagreement between the predicted mass function 
    error and the corresponding JK estimation in order to improve the latter. The $N_{JK}$ JK
    samples are constructed by subtracting haloes in JK cells of the size $V_{tot}/N_{JK}$ 
    from the total halo distribution.
    The number of haloes in the remaining JK sample
    is then given by $N^h_{JK} \equiv N^h_{tot} - N^h_{JKcell}$. The volume of a 
    JK sample is given by $V_{JK} \equiv V_{tot} - V_{JKcell}$.  
     From the definition of the number density, ($n\equiv N^h/V$), and the
     deviation from the mean over the total volume $\Delta\equiv n-\bar{n}$
     one can derive $\Delta_{JK} = - (N_{JK}-1)^{-1} \Delta(V_{JKcell})$.
     Note that this relation also holds for the number density contrast,
     $\delta_{JK} = - (N_{JK}-1)^{-1} \delta(V_{JKcell})$.
      Hence, subtracting an 
      overdense JK cell from the total volume generates a slightly underdense JK sample.
      This result leads to a relation between the variances of the number 
      density, $\sigma^2 \equiv \langle \Delta^2 \rangle$, in the JK cells, and the
      corresponding variance for the JK samples,
     
      \begin{equation}
        \sigma^2(V_{JKcell}) =
        (N_{JK}-1)^2 \langle \Delta_{JK}^2 \rangle.  
      \label{eq:JK_var}
      \end{equation}
      As for equation (\ref{eq:def_covar}) $\langle \ldots\rangle$ denotes the ensemble average.
       The variance of the JK samples is therefore simply related to the variance at the scales of the JK cells.
      Note that $\langle \Delta_{JK}^2 \rangle$ is not 
      the variance at the scale of the JK sample volume, $\sigma^2(V_{JK})$, since the JK 
      samples are not independent from each other.

\begin{figure}
      	\centerline{\includegraphics[width=72mm,angle=270]{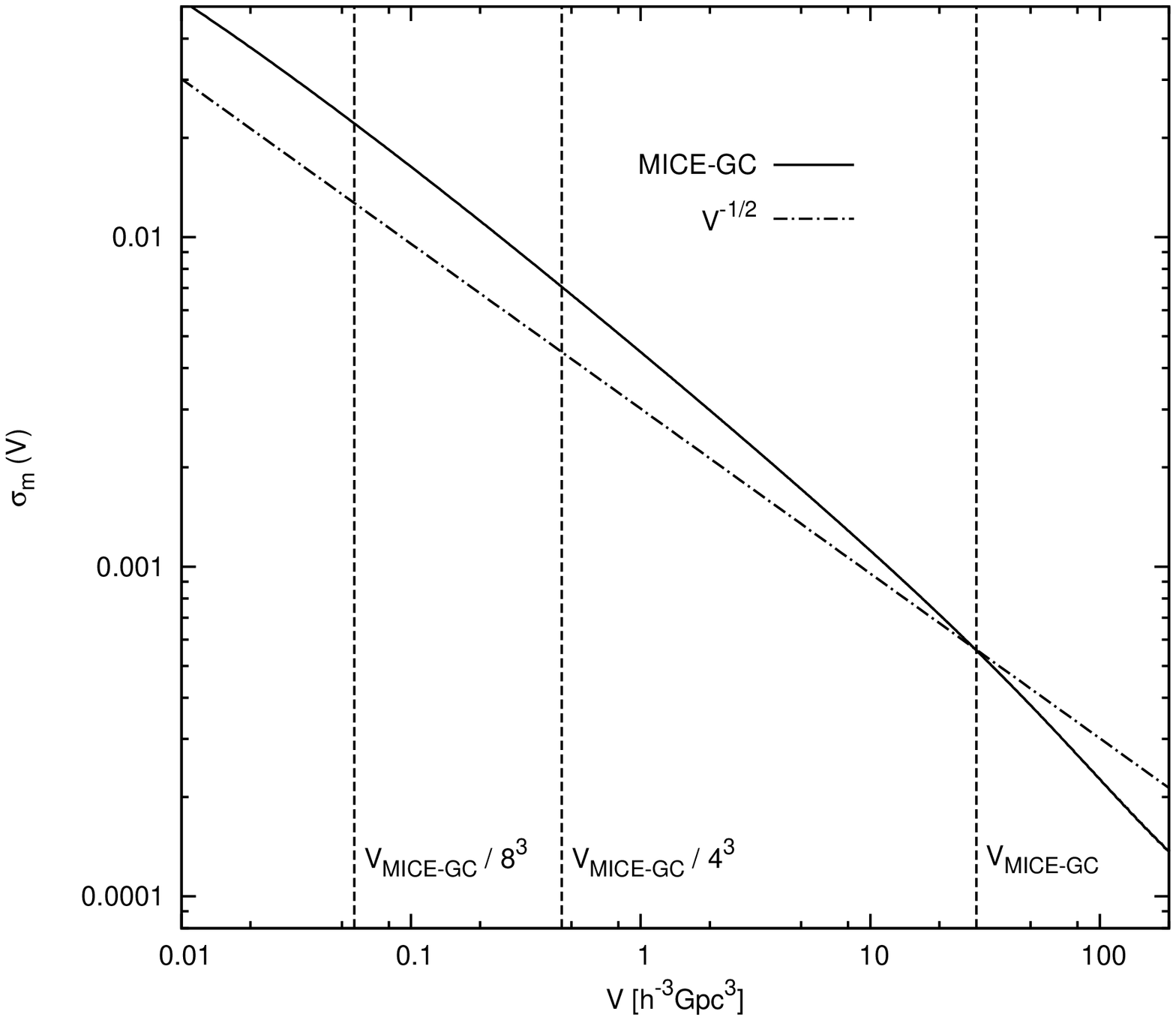}}
      	\caption{The standard deviation of matter fluctuations, $\sigma_m$, predicted from the linear
      	MICE-GC power spectrum via equation (\ref{eq:sigma}) as a function of the volume, $V$, of the spherical
      	top hat smoothing window (solid line). The dash-dotted line shows the $\sigma_m(V)$ relation
      	which corresponds to the standard JK estimator (with an arbitrary normalisation, chosen to
      	coincidence with the predictions at the MICE-GC simulation volume $V$).
      	The volumes of the simulation box and the JK cells are shown as vertical 
      	dashed lines.
      	}
      	\label{fig:sigma-vol}
      	\end{figure}

      From the linear bias model 
      we assume that the variance of the halo number density results 
      from shot-noise ($\sigma^2_{sn}$) and sampling variance ($\sigma^2_{s}$),
      as explained in Section \ref{sec:cov_prediction}.
      The latter contribution to the total variance of the JK cells, measured via
      equation (\ref{eq:JK_var}), is therefore given by
      \begin{equation}
      \sigma^2_s(V_{JKcell}) =  (N_{JK}-1)^2  \langle\Delta_{JK}^2\rangle - \sigma^2_{sn}(V_{JKcell}).
      \label{eq:JK_sampvar}
      \end{equation}
      Since $n/V_{JKcell}= N_{JK} n/V_{tot}$, the shot-noise for JK cells is related to the shot-noise
      of the whole box as $\sigma^2_{sn}(V_{JKcell}) = N_{JK} \sigma^2_{sn}(V_{tot})$.      
     To obtain the sampling variance at the scale of the simulation box,
     $\sigma^2_s(V_{tot})$, we multiply $\sigma^2_s(V_{JKcell})$
     with a rescaling factor
     \begin{equation}
        r_{\sigma} \equiv \sigma^2_{s}(V_{tot}) / 
     \sigma^2_{s}(V_{JKcell}),
      \label{eq:r_sig}
      \end{equation}
     which can be predicted from the linear matter power
     spectrum. This prediction is based on the assumption that,
     at large smoothing scales, the sampling variance of the halo number
     density is related to the dark matter variance by the linear 
     bias factor, $\sigma^2_{s} = b_1\sigma^2_{m}$. Since
     $b_1$ is constant at large scales (see Fig. \ref{fig:b1_2pccfits}),
     it cancels out in the rescaling factor, hence $r_{\sigma}^{h} = r_{\sigma}^{m}= r_{\sigma}$.
     The prediction is then based on $\sigma_{m}(V)$, computed from the linear
     matter power spectrum via equation (\ref{eq:sigma}).      
      We can now write the expression for the sampling variance of the simulation box,
      based on equation (\ref{eq:JK_sampvar}) in the general case of the covariance
      
      \begin{equation}
       C_{ij}^{s}(V_{tot}) = r_{\sigma} [(N_{JK}-1)^2 \langle\Delta_i\Delta_j\rangle - 
       N_{JK}C_{ij}^{sn}(V_{tot})].
      \label{eq:JK_sampling_var_box}
      \end{equation}
      Note that we have now dropped the index $JK$ in $\Delta$ for simplicity and to be 
      consistent with equation (\ref{eq:def_JKcovar}) for the standard JK estimator.
      As in the latter equation the lower indices refer to the mass bins $i$ and $j$.
      With the Poisson shot-noise term from equation (\ref{eq:covar_prediction_shot-noise}),
      the total covariance is then given by equation (\ref{eq:covar_tot}) as
      $C_{ij}=C_{ij}^{s}+C_{ij}^{sn}$. The resulting expression constitutes a new error estimation
      for the mass function which combines direct measurements of sampling variance
      via the JK sampling with predictions for the rescaling factor and 
      the shot-noise.
      This new estimator can be written more explicitly as
       \begin{equation}
        C^{new JK}_{ij} =
        r_{\sigma} (N_{JK}-1)^2 \langle \Delta_i \Delta_j \rangle + \delta_{ij}\frac{\sqrt{n_i n_j}}{V_{tot}} 
        (1-r_{\sigma} N_{JK}).
      \label{eq:new_jkcovar}
      \end{equation}
      As before the diagonal elements correspond to the variance, $\sigma_i^2=C_{ii}$.
     Note that for Poisson shot-noise dominated errors the sampling variance
       can be approximated as $C_{ij}^{s}(V_{tot}) \simeq 0$
      and the new estimator reduces to the shot-noise term,
      $ C^{new JK}_{ij} \simeq \delta_{ij}\frac{\sqrt{n_i n_j}}{V_{tot}}$. In 
      this case we derive from equation (\ref{eq:JK_sampling_var_box}) that
      $(N_{JK}-1)^2 \langle\Delta_i\Delta_j\rangle \simeq N_{JK}C_{ij}^{sn}(V_{tot})$.
      For large numbers of JK samples ($N_{JK} \simeq N_{JK}-1$)
      this expression is equivalent to 
      $(N_{JK}-1) \langle\Delta_i\Delta_j\rangle \simeq \delta_{ij}\frac{\sqrt{n_i 
      n_j}}{V_{tot}}$. The left hand side of this relation is the standard JK 
      estimator.
      This consideration explains the good agreement between standard JK estimator with the
      improved JK estimator and the predictions at high masses, where the errors and the covariance are 
      shot-noise dominated (Fig. \ref{fig:nofm_errcomp}, \ref{fig:covar} and \ref{fig:covar_diff}).
      In the low mass range our new method is in much better agreement with the predictions than the
      standard JK error estimator (\ref{eq:def_JKcovar}). This can be understood with the following consideration.
      For large numbers of JK samples ($N_{JK}\simeq N_{JK}-1$) the new estimator 
      corresponds to the
      standard JK estimator if
      $r_{\sigma} \equiv \sigma_m^2(V_{tot})/\sigma_m^2(V_{JK cell}) =  1/N_{JK} $.
      Since $V_{tot} = N_{JK} V_{JKcell}$, this condition is equivalent to
      $V_{tot} \sigma_m^2(V_{tot}) = V_{JK cell} \sigma_m^2(V_{JK cell})$. The 
      JK approach can therefore be described as the assumption that
      $\sigma_m(V)\sim V^{-1/2}$ for large $N_{JK}$. We show $\sigma_m(V)$, computed 
      from the linear power spectrum via equation (\ref{eq:sigma}) in Fig. 
      \ref{fig:sigma-vol}. The JK assumption is in a clear disagreement with the 
      prediction which causes a too high $r_{\sigma}$ and therefore an 
      overestimation of sampling variances at the scale of the simulation box for the standard JK assumption.
      
      The advantage of the new JK estimation with respect to the prediction is that it does 
      not require knowledge of the halo bias. Furthermore, this new approach 
      is independent of the normalisation of the power spectrum as it 
      cancels out in the rescaling factor (equation (\ref{eq:r_sig})).
      However, the large scale power spectrum still needs to 
      be known for accurate rescaling of the sampling variance via $\sigma_m(V)$. For simulations the
      linear power spectrum is given. In this case the new method can be used instead of running 
      several realisations for deriving mass function errors and covariances.
      In observations the large scale power spectrum can only be
      assumed. However, with such an assumption the accuracy of the error estimation 
      can still be improved with respect to the standard JK method, which also
      implies the strong assumption of $\sigma(V)\sim V^{-1/2}$.
      The advantage of the new JK estimation with respect to using independent 
      subvolumes for the error estimation is that the JK samples cover larger 
      volumes with larger average numbers of massive haloes. The covariance
      between the low- and high mass end of the mass function is therefore 
      better sampled by JK samples than subvolumes.
      We employ our new method for the error and covariance estimation using $N_{JK} = 8^3$ samples.
      
\section{PBS bias predictions}\label{app:pbs_bias}

\begin{figure*}
      \centerline{\includegraphics[width=120mm,angle=270]{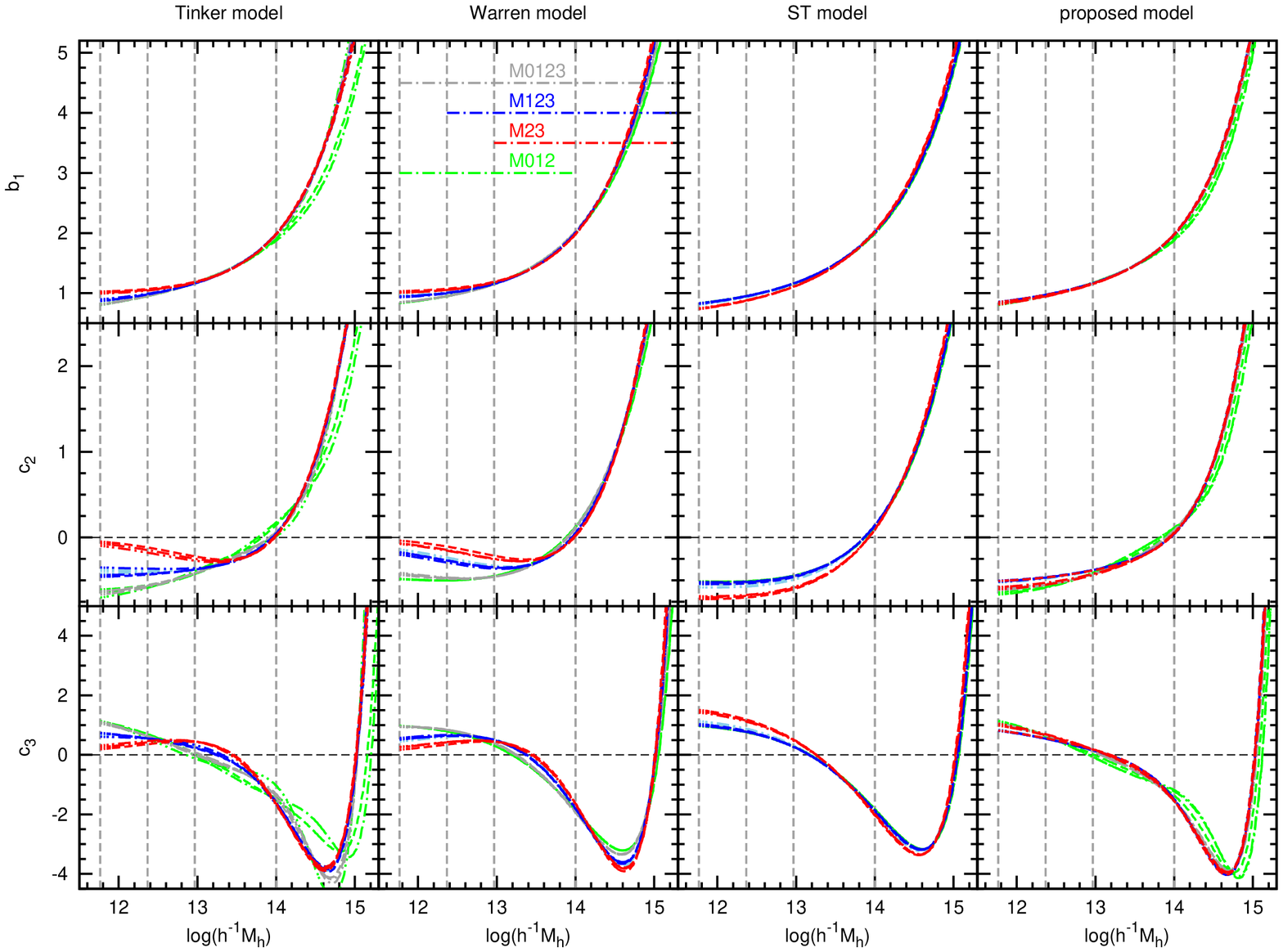}}
      \caption{Same as Fig. \ref{fig:b1c2c3_pbs}, but for all mass function models
	analysed in this work. In addition we show bias predictions based on mass 
	function fits over the whole MICE-GC mass range (M0123). For predictions from fits over 
	the mass range M123 we show in addition results derived without taking the covariance in the
	measurements between different mass function bins into account as light blue lines.
	Note that these lines are covered by other results in most cases.}.
      	\label{fig:b1c2c3_pbs_allmodels}
      	\end{figure*}

We show in Fig. \ref{fig:b1c2c3_pbs_allmodels} the PBS bias predictions based on the mass function models
studied in this analysis. The different predictions are based on fits over the four mass ranges
M0123, M123, M23 and M012, defined in Table \ref{table:mass_ranges}. The figure is analogous
to Figure \ref{fig:b1c2c3_pbs}. We find that the linear bias parameter $b_1$ is less sensitive to the mass function
model and the mass function fitting range than the non-linear bias parameters á$c_2\equiv b_2/b_1$ and $c_3\equiv b_3/b_1$.
I addition to Figure \ref{fig:b1c2c3_pbs} we show that the bias predictions become unstable
when the low mass sample, M1, is included in the analysis. Furthermore we show bias predictions,
based on mass function fits over the range M123, which were  derived neglecting the
covariance between different mass function bins. We find that the for this example the mass function
covariance has a smaller impact on the bias prediction than the choice of the mass function model,
or the mass function fitting range. We expect the impact of the mass function covariance to increase,
when low mass samples are included in the fit. However, the low mass range is hard to access for
analysis of halo abundance in the MICE-GC, due to resolution effects (see Section \ref{sec:mass_function_fits}).

